\documentclass[5p]{elsarticle}

\usepackage[colorlinks=true]{hyperref}
\usepackage[english]{babel}
\usepackage[utf8]{inputenc}

\journal{Journal of Theoretical Biology}

\biboptions{sort&compress}

\usepackage{subfigure}
\usepackage{amsmath}
\usepackage{color}
\usepackage{multirow}
\usepackage{amssymb}
\usepackage{multicol}
\begin{document}

\begin{frontmatter}

\title{An intramembranous ossification model for the \textit{in-silico} analysis of bone tissue formation in tooth extraction sites \\
{\small Accepted manuscript for the scientific article with the same name, which was published in the Journal of Theoretical Biology (2016).
\url{https://doi.org/10.1016/j.jtbi.2016.04.023}}}


\author[unalBIO,GATAcorp]{Jennifer Paola Corredor-Gómez\corref{mycorrespondingauthor}}
\cortext[mycorrespondingauthor]{Corresponding author \\
\textit{Carrera 30 45-03, Edificio 407, Oficina 213}.}
\ead{jpcorredorg@unal.edu.co}

\author[unalBIO,GATAcorp]{Andrés Mauricio Rueda-Ramírez}
\author[unalBIO,GATAcorp]{Miguel Alejandro Gamboa-Márquez}
\author[unalBIO,unalOD]{Carolina Torres-Rodríguez}
\author[unalBIO]{Carlos Julio Cortés-Rodríguez}

\address[unalBIO]{Research Group on Biomechanics (GIBM-UN). Mechanical and Mechatronics Engineering.\\ Universidad Nacional de Colombia, Sede Bogotá.}
\address[GATAcorp]{Advanced Technologies Analysis Group (GATAcorp) }
\address[unalOD]{Department of Dentistry. Universidad Nacional de Colombia, Sede Bogotá.}

\begin{abstract}
The accurate modeling of biological processes allows to predict the spatiotemporal behavior of living tissues by computer-aided (\textit{in-silico}) testing, a useful tool for the development of medical strategies, avoiding the expenses and potential ethical implications of \textit{in-vivo} experimentation. A model for bone healing in mouth would be useful for selecting proper surgical techniques in dental procedures. In this paper, the formulation and implementation of a model for Intramembranous Ossification is presented aiming to describe the complex process of bone tissue formation in tooth extraction sites. The model consists in a mathematical description of the mechanisms in which different types of cells interact, synthesize and degrade extra-cellular matrices under the influence of biochemical factors. Special attention is given to angiogenesis, oxygen-dependent effects and growth factor-induced apoptosis of fibroblasts. Furthermore, considering the depth-dependent vascularization of mandibular bone and its influence on bone healing, a functional description of the cell distribution on the severed periodontal ligament (PDL) is proposed. The developed model was implemented using the finite element method (FEM) and successfully validated by simulating an animal \textit{in-vivo} experiment on dogs reported in the literature. A good fit between model outcome and experimental data was obtained with a mean absolute error of 3.04\%. The mathematical framework presented here may represent an important tool for the design of future \textit{in-vitro} and \textit{in-vivo} tests, as well as a precedent for future \textit{in-silico} studies on osseointegration and mechanobiology.
\end{abstract}


\end{frontmatter}

{
\scriptsize 
\let\thefootnote\relax\footnotetext{\\ \textbf{Abbreviations}:}
\let\thefootnote\relax\footnotetext{BC: Blood clot}
\let\thefootnote\relax\footnotetext{H-D: Hypoxia-dependent}
\let\thefootnote\relax\footnotetext{IOM: Intramembranous ossification model}
}

\section{Introduction}
A deep understanding of the complex biological and biochemical processes in bone's metabolism and the consequent capacity to predict its behavior has contributed to the development of novel medical techniques such as the localized administration of growth factors for assisted bone healing \cite{luginbuehl2004localized}, the ultrasound-mediated bone regeneration \cite{mitragotri2005healing}, the careful and personalized selection of bone grafts, the development of scaffolds for regenerative medicine, among many others that still have to be considered. Such an understanding has been traditionally obtained from \textit{in-vitro} and \textit{in-vivo} experimentation involving high costs at several levels. However, the latest advancements in computational modeling and computer power have boosted a new field of study that can shed light on such matters: \textit{in-silico} testing. 

It is possible to put forward mathematical models to describe the behavior of living tissues \cite{doblare2004modelling}. The in silico implementation of these models has a potential role in research and development of treatments, implants design and even in the prognosis of bone pathologies \cite{isaksson2012recent}, significantly reducing the need for traditional testing techniques. Several mathematical models have been developed in order to understand the fundamentals of cell and tissue biology, and  the behavior of biochemical factors in bone \cite{lacroix2002mechano,gomez2005influence,bailon2001mathematical,geris2010connecting,garcia2007computational,pivonka2010mathematical, komarova2003mathematical,ribeiro2015silico,geris2008angiogenesis,peiffer2011hybrid,carlier2012mosaic}. However, while most of the efforts have focused on studying regeneration and remodeling in long bones, to the authors' knowledge, few \textit{in-silico} works emphasize bone formation in the mandible, specifically regarding to intramembranous ossification (IO).

This paper presents the formulation of a mathematical model for predicting the spatiotemporal behavior of bone healing via IO. The model was validated by comparing its results with an \textit{in-vivo} study of bone regeneration in tooth extraction sites. The proposed model builds on previous mathematical descriptions of fracture healing in long bones and incorporates modifications to account for the particular phenomena that take place in the mandibular bone. It also takes into account the findings by recent studies on the topic and introduces new biological and numerical considerations such as the growth factor-mediated apoptosis and a detailed description of the alveolar boundary conditions.\\

Our mathematical model can be used to predict the behavior of particular types of cells, the formation and degradation of important tissues and the action of growth factors during intramembranous bone healing, even in geometries that differ from the one used for validation. This makes possible to predict the events following a clinical intervention and to analyze some situations of compromised and assisted healing.\\

Due to the immense complexity of bone metabolism, and the lack of quantitative data and computational capacity \cite{zhang2012role}, it is currently impossible to reproduce all the interacting biological, biochemical and mechanical processes by means of \textit{in-silico} testing. This problem was tackled by making some simplifications and assumptions that helped the development of the current model. Furthermore, since a detailed histological examination of the alveolar bone regeneration in humans has many inherent difficulties, the validation of the model was performed by comparing the simulations outcome with an \textit{in-vivo} study conducted in dogs as it is normally done.

\section{Theoretical background: non-pathological bone healing} \label{bonehealing}
Bone tissue formation (also termed as ossification or osteogenesis) following injury is a complex process that involves the action of several cells and molecules. The different types of cells migrate, proliferate, differentiate, interact, synthesize and degrade extra-cellular matrices (ECM), proteins and enzymes under the influence of mechanical stimuli \cite{gerstenfeld2003fracture} and a variety of regulatory molecules. These molecules include insulin-like growth factors (IGFs), fibroblast growth factors (FGFs), platelet derived growth factors (PDGFs), transforming growth factors (TGFs), and bone morphogenetic proteins (BMPs), among others \cite{barnes1999growth}. This process has been described by various authors as a recapitulation of the different embryological processes of bone formation: inflammation, repair and remodeling  \cite{sikavitsas2001biomaterials,ferguson1999does,vortkamp1998recapitulation}.\\

During bone formation, there are two distinct and interactive responses: endochondral and intramembranous ossification \cite{gerstenfeld2003fracture}. Endochondral ossification is characterized by chondrogenic growth factors and chondrocyte activity that allow the existence of cartilage; on the contrary, these aspects are absent in IO. Although the wound healing process in long bones (femur, tibia, ulna, etc.) has been clearly identified to occur via both responses \cite{reed2003vascularity,kokubu2003development,harrison2003controlled}, there has been an active discussion about which response takes place during bone formation in the mandible \cite{frommer1971contribution}.\\

Both intramembranous and endochondral ossification have been reported to occur in the mandible during fetal development  \cite{shibata2014distribution}. Nevertheless, many studies report absence of cartilage during bone regeneration, as will be seen later. De facto, three ossification centers operate in the mandible during fetal development: two predominant are intramembranous, whereas the third is an endochondral center within the Meckel's cartilage \cite{frommer1971contribution}. Endochondral ossification occurs only where the cartilage bar is near to the primary intramembranous centers; the process is transient and dimmed by the IO. After the cartilage is calcified, it is quickly invaded and surrounded by membrane bone and the endochondral osteogenic zones rapidly lose their identity \cite{frommer1971contribution}.\\

One of the most common processes of bone healing in the mandible is the one that occurs inside an alveolus left by the extraction of teeth. It has been subject of multiple studies, both in animal and human specimens \cite{cardaropoli2003dynamics,araujo2005dimensional,kuboki1988time,
simpson1960experimental,amler1969time,christopher1941histological}. All authors describe the healing process in the extraction socket in the following sequence (note the absence of cartilage):
\begin{enumerate}
\item Formation and maturation of a blood clot (BC), which serves as initial construct for the healing process.
\item Replacement of the BC with a provisional matrix (PCT) synthesized by fibroblasts, which soon becomes the main construct for vascularization and cell migration.
\item Replacement of the PCT with newly formed woven bone.
\item Bone mineralization.
\item Bone remodeling.
\end{enumerate}

At the time of tooth extraction, blood vessels are damaged resulting in bleeding and the consequent formation of a BC. The platelets contained in the blood release PDGF and TGF-$\beta$, together with vasoactive factors in a phenomenon called platelet degranulation \cite{davies2003understanding}, which concludes with vasoconstriction and the generation of a fibrin network \cite{davies2003understanding}. The newly formed network serves as a scaffold for the migration of fibroblasts and mesenchymal stem cells (MSCs) that are attracted by both PDGF and TGF-$\beta$ \cite{chandrasekhar1996modulation} and are originally immersed inside the bone marrow stroma \cite{hernandez2006physiological}. BC is gradually degraded by the action of macrophages and plasmin (a protein contained in plasma), contributing to the migration of MSCs and fibroblasts \cite{li2003plasminogen}. This initiates fibroplasia: the replacement of the blood clot with PCT and granulation tissue (GT), collagen-rich ECMs synthesized by fibroblasts which serve as a construct for angiogenesis \cite{aukhil2000biology}. While the fibrin clot forms within minutes \cite{vanegas2011finite}, its degradation can take up to 7 days \cite{cardaropoli2003dynamics}.\\

Angiogenesis is initiated from the severed blood vessel ends and is mainly mediated by the presence of vascular-endothelial growth factors (VEGFs) \cite{ferrara1997biology}, which are endothelial cell mitogens and chemoattractants. These factors are produced in early stages of bone healing by cells in hypoxia conditions \cite{pugh2003regulation,maes2012hypoxia,komatsu2004activation}. The resulting formation of vasculature is one of the primary driving mechanisms that facilitate the intramembranous bone formation \cite{gerstenfeld2003fracture,hirao2006oxygen}.\\

After the blood supply has been restored, a new phase begins. PCT is replaced by woven bone formed by osteoblasts, which differentiate from MSCs going through distinct intermediate stages (osteogenic cells) under the influence of several developmental signals. Among these are the Hedgehog, Wnt, Notch, BMP and FGF signaling pathways \cite{long2012building}. Osteogenic cells are chemotactically attracted by PDGF and TGF-$\beta$ into the PCT \cite{chandrasekhar1996modulation} and afterwards mature osteoblasts synthesize a matrix consisting of 90\% collagen proteins and 10\% non-collagen proteins \cite{anselme2000osteoblast,sikavitsas2001biomaterials}. This matrix is then mineralized and remodeled, completing the regeneration process.

\section{Materials and Methods}

\subsection{Mathematical model for bone healing} \label{model}
The mathematical framework used in this study builds on the bioregulatory model developed by Geris \textit{et al.} \cite{geris2008angiogenesis}, which is based upon the work of Bailón-Plaza \textit{et al.} \cite{bailon2001mathematical} and Olsen \textit{et al.} \cite{olsen1997mathematical}. These models describe the biological phenomena that take place during fracture healing in long bones by modeling the concentration of several cell types in number of cells per unit volume ($c_i$), the density of different ECMs ($m_j$) and the concentration of various growth factors ($g_k$), both in mass per unit volume.\\

Later models by Peiffer \textit{et al.} \cite{peiffer2011hybrid} and Carlier \textit{et al.} \cite{carlier2012mosaic,carlier2015oxygen} were also reviewed and taken into account for the formulation of the intramembranous ossification model (IOM). Nevertheless, the discrete approximation of the blood vessels formation, which these latter models introduced, has not been adopted, since the results of the computational investigation by Geris \textit{et al.} \cite{geris2008angiogenesis} show that the continuous approximation is accurate enough for the description of the angiogenesis influence on bone healing. The mathematical formulation consists in three different sets of equations.

For cell concentration:

\begin{small}
\begin{equation} \label{eq:cells}
\begin{split}
\frac{{\partial c_i }}{{\partial t}} = \nabla  \cdot \left[ {\underbrace {D_i \nabla c_i }_{Diffusion} - \underbrace {C_{i,CT} c_i \nabla {g_i } }_{Chemotaxis} - \underbrace {C_{i,HT} c_i \nabla m}_{Haptotaxis}} \right] \\ 
+ \underbrace {A_i c_i \left[ {1 - \alpha _i c_i } \right]}_{Proliferation} + \underbrace {\sum\limits_{l\ne i} {F_{li} c_l } }_{\rightarrow Differentiation} \\
- \underbrace {\sum\limits_{n\neq i} {F_{in} c_i } }_{Differentiation \rightarrow } - \underbrace {d_i c_i }_{Apoptosis}
\end{split}
\end{equation}

For ECM density:
\begin{equation}
\label{eq:matrices}
\frac{{\partial m_j }}{{\partial t}} = \underbrace {P_{js} \left( {1 - \kappa _j m_j } \right)c_j }_{Synthesis} - \underbrace {Q_j m_j c_s \prod\limits_r {m_r } }_{Resorption}
\end{equation}

For growth factor concentration:
\begin{equation} \label{eq:gf}
\frac{{\partial g_k }}{{\partial t}} = \underbrace {\nabla  \cdot \left[ {D_{gk} \nabla g_k } \right]}_{Diffusion} + \underbrace {\sum\limits_p {E_{gkp} c_p } }_{Production}   - \underbrace {\sum\limits_t{d_{gkt} g_k c_t }}_{Consumption} - \underbrace {d_{gk} g_k }_{Decay}
\end{equation}
\end{small}

The first term of equation \ref{eq:cells} describes the migration of each cell type as a combination of diffusion, chemotaxis and haptotaxis. The diffusion term accounts for the random movement of cells, while the chemotactic and haptotactic terms represent the directed motion. The cell proliferation rate is considered proportional to the cell concentration up to a saturation value ($\alpha_i^{-1}$). The $\rightarrow$\textit{Differentiation} (differentiation from) term represents the differentiation into $c_i$ from $c_l$, whereas the \textit{Differentiation}$\rightarrow$  (differentiation towards) term is associated with the decrease of $c_i$ because of its differentiation into another cell type. The last term refers to cell apoptosis. The first term of equation \ref{eq:matrices} refers to the ECM synthesis by a specific type of cell (up to a maximum value $\kappa_j^{-1}$) and the second to its consumption, which can be associated with the formation of another ECM ($m_r$) or with the presence of certain type of cells ($c_s$). Equation \ref{eq:gf} describes the changes in the concentration of the growth factor $g_k$ as a result of its diffusion, its production by different cells ($c_p$), its natural decay and its consumption by some type of cell ($c_t$).\\

\begin{figure} 
\centering
\includegraphics[width=1\linewidth]{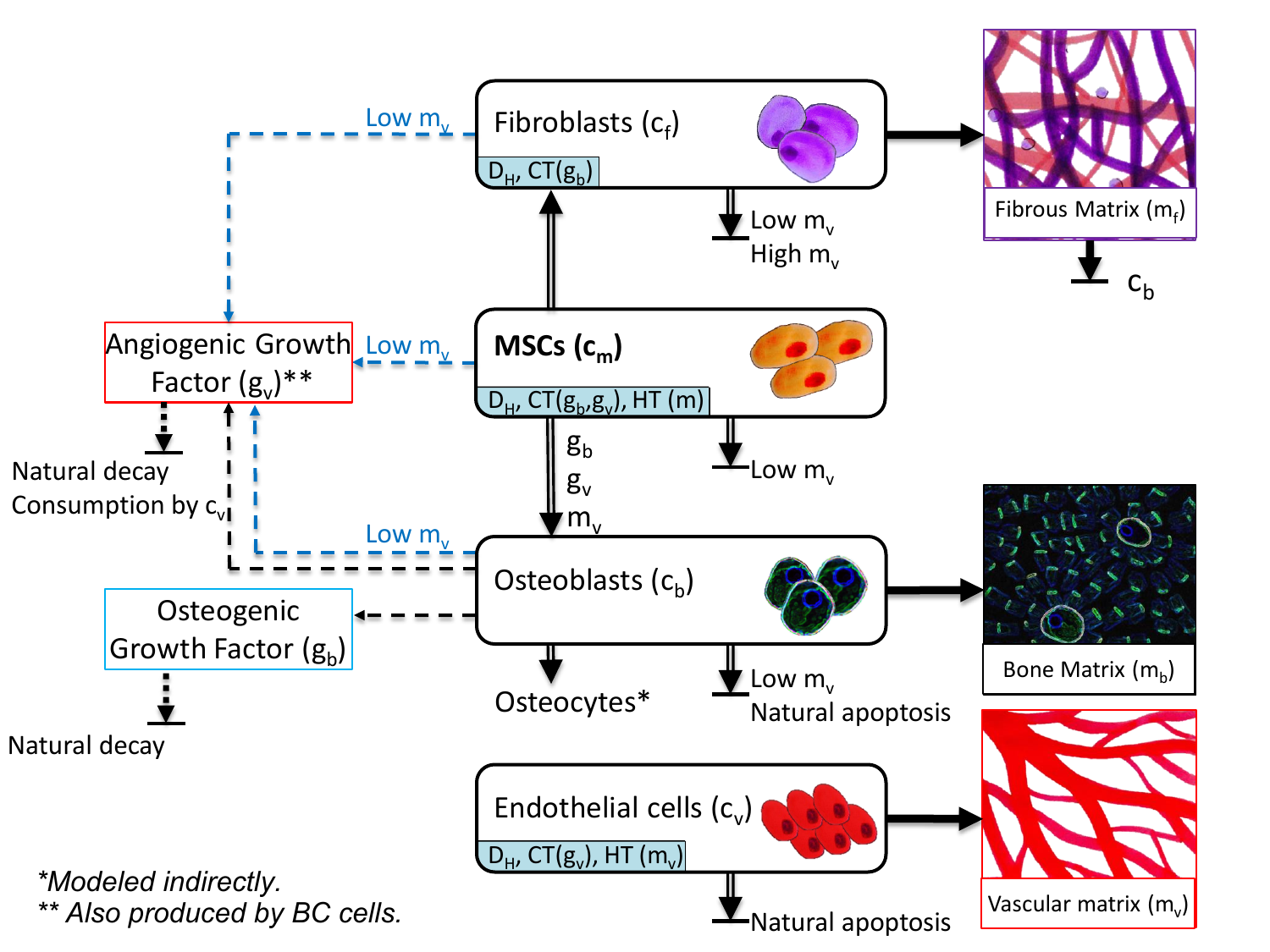}
\caption{Graphical representation of the mathematical model. The inner blue boxes indicate the relevant modes of motility simulated for each cell: $D_H$ stands for haptokinetic diffusion, CT for chemotaxis (chemo-attractant in brackets) and HT for haptotaxis (substrate in brackets). Dashed arrows indicate production of growth factors, double line arrows indicate differentiation and continuous line arrows indicate matrix synthesis. Blue dashed arrows refer to the H-D production of angiogenic growth factor ($g_v$). Arrows ending in a perpendicular line (\underline{$\Downarrow$}) indicate: decay or consumption of growth factors (dashed arrows); apoptosis (double arrows); tissue resorption or degradation (continuous arrows). \textit{Inspired on the schematic representation by Geris \textit{et al.}} \cite{geris2008angiogenesis}.} 

\label{fig:graphicalmodel}
\end{figure}

A schematic representation of the IOM is shown in Figure \ref{fig:graphicalmodel}, which is mathematically described by the following system of 9 coupled partial differential equations (PDE):\\

\begin{small}
\begin{align}
\frac{{\partial c_m }}{{\partial t}} &= \begin{aligned}[t]
										& \nabla  \cdot \left[ {D_m \nabla c_m  - C_{m,CT} c_m \nabla \left( {g_b  + g_v } \right) }\right] \\
										& -  \nabla  \cdot \left[ {C_{m,HT} c_m \nabla m} \right]+ A_m c_m \left[ {1 - \alpha _m c_m } \right]  \\
										& - F_{mb} c_m  - F_{mf} c_m  -d_m c_m
										\end{aligned} \label{eq:cm}\\
\frac{{\partial c_f }}{{\partial t}} &= \begin{aligned}[t]
										&\nabla  \cdot \left[ {D_f \nabla c_f  - C_{f,CT} c_f \nabla g_b } \right] \\
										&+ A_f c_f \left[ {1 - \alpha _f c_f } \right] + F_{mf} c_m  - d_f c_f 
										\end{aligned}\\
\frac{{\partial c_b }}{{\partial t}} &= A_b c_b \left[ {1 - \alpha _b c_b } \right] + F_{mb} c_m  - F_{bo} c_b  - d_b c_b \\
\frac{{\partial c_v }}{{\partial t}} &= \begin{aligned}[t]
										& \nabla  \cdot \left[ {D_v \nabla c_v  - C_{v,CT} c_v \nabla g_v  - C_{v,HT} c_v \nabla m_v } \right] \label{eq:cv}\\
										& + A_v c_v \left[ {1 - \alpha _v c_v } \right] - d_v c_v 
										\end{aligned}\\
\frac{{\partial m_f }}{{\partial t}} &= P_{fs} \left( {1 - \kappa _f m_f } \right)c_f  - Q_f m_f c_b \label{eq:mf}\\
\frac{{\partial m_b }}{{\partial t}} &= P_{bs} \left( {1 - \kappa _b m_b } \right)c_b \label{eq:mb}\\
\frac{{\partial m_v }}{{\partial t}} &= P_{vs} \left( {1 - \kappa _v m_v } \right)c_v \label{eq:mv}\\
\frac{{\partial g_b }}{{\partial t}} &= \nabla  \cdot \left[ {D_{gb} \nabla g_b } \right] + E_{gbb} c_b  - d_{gb} g_b \\
\frac{{\partial g_v }}{{\partial t}} &= \begin{aligned}[t]
										&\nabla  \cdot \left[ {D_{gv} \nabla g_v } \right] + E_{clot} m_f + E_{gvm} c_m  + E_{gvf} c_f  \\
										&+ E_{gvb} c_b  - g_v \left( {d_{gv}  + d_{gvc} c_v } \right)
										\end{aligned} \label{eq:gv}
\end{align}
\end{small}

\begin{table}[htbp]
  \centering
  \caption{Description of variables for each biological set.}
 \begin{small}
 \begin{tabular}{lcl}
    \hline
    \multicolumn{1}{c}{Biological set} & Variable & \multicolumn{1}{c}{Definition}\\
    \hline
    \multirow{4}[0]{*}{Cells} &$c_m$& Mesenchymal Stem Cells\\
          & $c_f$ & Fibroblasts \\
          & $c_b$ & Osteoblast \\
          & $c_v$ & Endothelial Cells \\
    \hline
    \multirow{3}[0]{*}{Tissues (ECMs)} & $m_f$ & Fibrous matrix \\
          & $m_b$ & Bone matrix \\
          & $m_v$ & Vascular matrix \\
    \hline         
    \multirow{2}[0]{*}{Growth factors} & $g_b$ & Osteogenic growth factor \\
          & $g_b$ & Vascular growth factor \\
    \hline         
    \end{tabular}%
 \end{small}
  \label{tab:defvar}%
\end{table}%

Where the four types of cells are MSCs ($c_m$), fibroblasts ($c_f$), osteoblasts ($c_b$) and endothelial cells ($c_v$); the three different ECMs are fibrous matrix ($m_f$), bone matrix ($m_b$) and vascular matrix ($m_v$); and the two different generic growth factors are osteogenic growth factor ($g_b$) and angiogenic growth factor ($g_v$) (Table \ref{tab:defvar}).\\

Since the ossification response that takes place in tooth extraction sites is intramembranous, the equations describing chondrocyte concentration, density of cartilaginous ECM and concentration of chondrogenic growth factor were removed from the original set of equations. As in previously reviewed models by Geris \textit{et al.} \cite{geris2008angiogenesis,geris2010connecting}, Peiffer \textit{et al.} \cite{peiffer2011hybrid} and Carlier \textit{et al.} \cite{carlier2012mosaic,carlier2015oxygen}, the fibrous matrix corresponds to BC, GT and PCT. Likewise $c_m$ represent not only MSCs but also intermediate stages of differentiation. Additional changes to the original mathematical model proposed by Geris \textit{et al.} \cite{geris2008angiogenesis} will be discussed later in this article.\\

Note that chemotaxis of osteoblasts has been ignored here because it is negligible, as reported by Peiffer \textit{et al.} \cite{peiffer2011hybrid}. It is also noteworthy that in all the reviewed models, both the fibroblast apoptosis and the fibrous matrix resorption are dependent on the appearance of cartilage (either directly or indirectly, due to the hypertrophy of the chondrocytes). However, this dependency has not been adopted in the present work, since there is no chondrogenesis. This aspect is discussed more extensively in section \ref{results}. As proposed by Carlier \textit{et al.} \cite{carlier2015oxygen}, all but the endothelial cells produce angiogenic growth factor in hypoxic conditions. It is important to mention that all PDEs must be non-dimensionalized. The non-dimensional variables will be denoted with a tilde (see \ref{appendixA}).\\

Further modifications to the original model of 2008 were made for three main reasons. First, so that the FEM implementation be numerically more stable. Second, in order to adopt important modifications made in more recent publications \cite{peiffer2011hybrid,carlier2012mosaic,carlier2015oxygen}; and finally, to consider additional biological phenomena reported in the literature.\\

The conducted procedure to determine the parameters' values used in the model was made through the following steps:

\begin{enumerate}
\item Parameters from previous models were adopted.
\item An extensive literature review was made in order to update and verify the parameters' values. This data was found in different \textit{in-vivo} and \textit{in-vitro} studies, leading to a modification of some of the previously adopted parameters and also to an incorporation of new parameters.
\item The adopted parameters were non-dimensionalized. (See \ref{appendixA} for detail).
\item Since some of the parameters' values were not reported in the literature and \textit{in-vivo} or \textit{in-vitro} experiments could not be performed, an exhaustive sensitivity analysis was conducted in the interest of estimate the numerical values.
\end{enumerate}

In the following subsections, we present the most important modifications incorporated into the IOM.\\


\subsubsection{Modifications to improve the numerical stability}
In the FEM implementation, the small magnitudes of $\tilde m_v$ and $\tilde c_v$, together with the larger values of the other variables that result from the approach developed by Geris \textit{et al.} \cite{geris2008angiogenesis}, led to inconveniences regarding the condition number of the system matrices ($\tilde c_v=O (10^{-4})$, $\tilde m_v=O (10^{-7})$, $\tilde g_b=O (10)$, etc.). This problem was overcome by changing the non-dimensionalization, \textit{i.e.} choosing the characteristic concentration of endothelial cells as $c_{v,0}=100$ cells/ml (shown in more detail in \ref{appendixA}). This leads to a boundary value of $\tilde c_{v,bc}=1$ in the original long bone non-union model.\\

Taking into account that the resulting $\tilde c_{v,bc}$ is the same as in the model of Olsen \textit{et al.} \cite{olsen1997mathematical}, $m_{v,0}$ was also changed in order to adopt their approximation for the vascular matrix synthesis rate by endothelial cells ($\tilde P_{vs}$), as well as the ECM saturation value ($m_{v,max}=\kappa_v^{-1}$). Due to the new non-dimensionalization for the vascular matrix and the lack of quantitative data in the literature, $C_{v,HT}$ was redefined in the continuous model by adopting the $D_v/C_{v,HT}$ ratio in accordance with the parameters reported by Olsen.\\

The value of $\tilde \alpha_v$ that was derived from the data of Geris \textit{et al.} \cite{geris2008angiogenesis} resulted to be too small in comparison with the value of other variables, as it allowed the growth of the endothelial cell population up to $\tilde c_v \approx O(10^3)$ in some points inside the domain (data not shown).  In order to obtain homogeneous order of magnitude of the non-dimensionalized variables, the saturation value for the proliferation of endothelial cells was set to $\tilde \alpha_v=1$, which is consistent with the value reported by Olsen \textit{et al.} \cite{olsen1997mathematical}. \\

Another form of numerical instability like oscillatory behavior could occur due to the convective nature of the chemotactic and haptotactic terms. It imposes a problem for the stability of the Petrov-Galerkin discrete solutions. These instabilities were dealt with by incorporating the stabilization procedure for chemotaxis problems proposed by Strehl \textit{et al.} \cite{strehl2010flux,strehl2013positivity}.

\begin{figure} 
  \begin{minipage}[b]{0.450\linewidth}
    \centering
    \includegraphics[width=\linewidth]{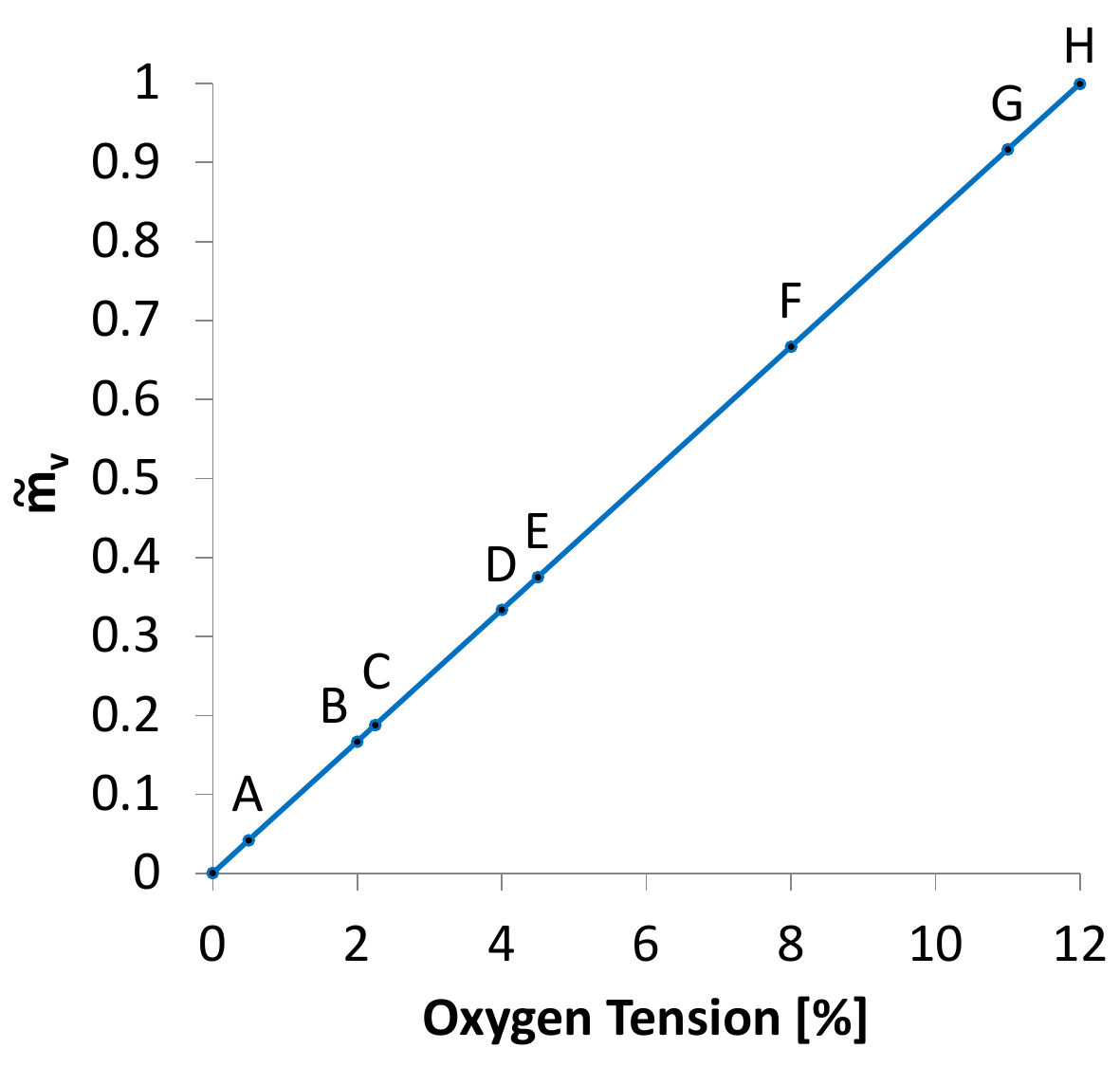} 
    \par\vspace{0pt}
  \end{minipage}%
  \begin{minipage}[b]{0.55\linewidth}
    \centering%
\begin{scriptsize}
    \begin{tabular}{cp{0.7\linewidth}} 
    \hline
   Label & Description\\
    \hline
    A & \textit{H-D} $c_m$ apoptosis  \\
    B & \textit{H-D} production of $g_v$ by $c_m$\\
      & \textit{H-D} $c_b$ apotosis \\
    C & \textit{H-D} $c_f$ apotosis \\
    D & \textit{H-D} production of $g_v$ by $c_b$\\
    E & \textit{H-D} production of $g_v$ by $c_f$\\
    F & $m_{v,min}$ for $c_m$ \textit{diff}$\rightarrow$ $c_b$ \\
    G & bFGF-dependent $c_f$ apoptosis* \\
    H & Maximum oxygen tension\\
    \hline
    \multicolumn{2}{l}{\textit{*New: numerically approximated}} \\
    \multicolumn{2}{l}{} \\
    \end{tabular}%
\end{scriptsize} 
    \par\vspace{0pt}
  \end{minipage}
\caption{Assumed functional relationship between oxygen tension and $\tilde m_v$. The processes are listed in the table according to the data reported by Carlier \textit{et al.} \cite{carlier2015oxygen}}
\label{fig:oxygen}
\end{figure}

\subsubsection{Incorporation of vascularization-dependent effects} \label{vascudependent}
Carlier \textit{et al.} \cite{carlier2015oxygen} made an exhaustive literature review about the influence of oxygen tension in most of the biological processes that occur during bone fracture healing. Their approach was adopted in the present work and adapted to the continuous angiogenesis under two assumptions: first, that the oxygen tension in each part of the domain is proportional to the vascularization ($\tilde m_v$); and second, that the maximum oxygen tension during bone healing (taken as 12\% as reported by Brighton \textit{et al.} \cite{brighton1972oxygen}) corresponds to the maximum value of $\tilde m_v$. The following oxygen-dependent processes were considered (see Figure \ref{fig:oxygen}):
\begin{enumerate}
\item Apoptosis due to hypoxia (labels A, B and C)\cite{grayson2007hypoxia,chae2001hypoxia}.
\item H-D production of angiogenic growth factor (labels B, D and E) \cite{fraisl2009regulation,carlier2015oxygen}.
\item Minimum oxygen tension for intramembranous ossification (label F) \cite{hirao2006oxygen}, taken as that required for osteoblast differentiation from MSCs \cite{carlier2015oxygen}. 
\end{enumerate}

In a recent study, Akasaka \textit{et al.} \cite{akasaka2010mechanisms} found that the mechanisms underlying fibroblast apoptosis during wound healing are closely related with the appearance of certain growth factors. They reported that the basic fibroblast growth factor bFGF (or FGF-2, as it is also known) promotes the apoptosis of injured tissue-derived fibroblasts that were previously treated with transforming growth factor TGF-$\beta$1. It is well known that endothelial cells express bFGF to produce autocrine effects \cite{schweigerer1987capillary,vlodavsky1987endothelial,seghezzi1998fibroblast} and that TGF-$\beta$1 is produced by platelets during the initial clot formation \cite{mehta2008platelet} \cite{marx2001platelet}. Hence, an increase in fibroblast apoptosis can be expected at high vascularizations.\\

Taking into account the reasons mentioned above, the current analysis introduces fibroblast apoptosis as an additional process that depends $\tilde m_v$. Because of the lack of available quantitative experimental data in the literature, the fibroblast apoptosis term was firstly adopted in the same way as Carlier \textit{et al.} described the apoptosis of chondrocytes due to high oxygen tensions \cite{carlier2015oxygen}. Nevertheless, the magnitude of the apoptosis rate was changed numerically, in order to reproduce the experimental data of Cardaropoli \cite{cardaropoli2003dynamics} (see section \ref{results} and label G in Figure \ref{fig:oxygen}). 

\subsubsection{Thresholding models} \label{thresholding}
Previous works used sixth-order Hill functions to model thresholds \cite{geris2008angiogenesis,bailon2001mathematical, peiffer2011hybrid,carlier2015oxygen}. In the case of fibroblast apoptosis, the biological phenomena discussed above is expressed mathematically by an equation of the form:\\
\begin{small}
\begin{equation} \label{eq:threshold}
\tilde d_f  =  \frac{{\tilde d_{f,hyp} \tilde H_{hyp}^6 }}{{\tilde H_{f,hyp}^6  + \tilde m_v^6 }} + \frac{{\tilde d_{f,bFGF} \tilde m_v^6 }}{{\tilde H_{f,bFGF}^6  + \tilde m_v^6 }}
\end{equation}
\end{small}

\begin{figure} 
\begin{center}
    \includegraphics[width=0.65\linewidth]{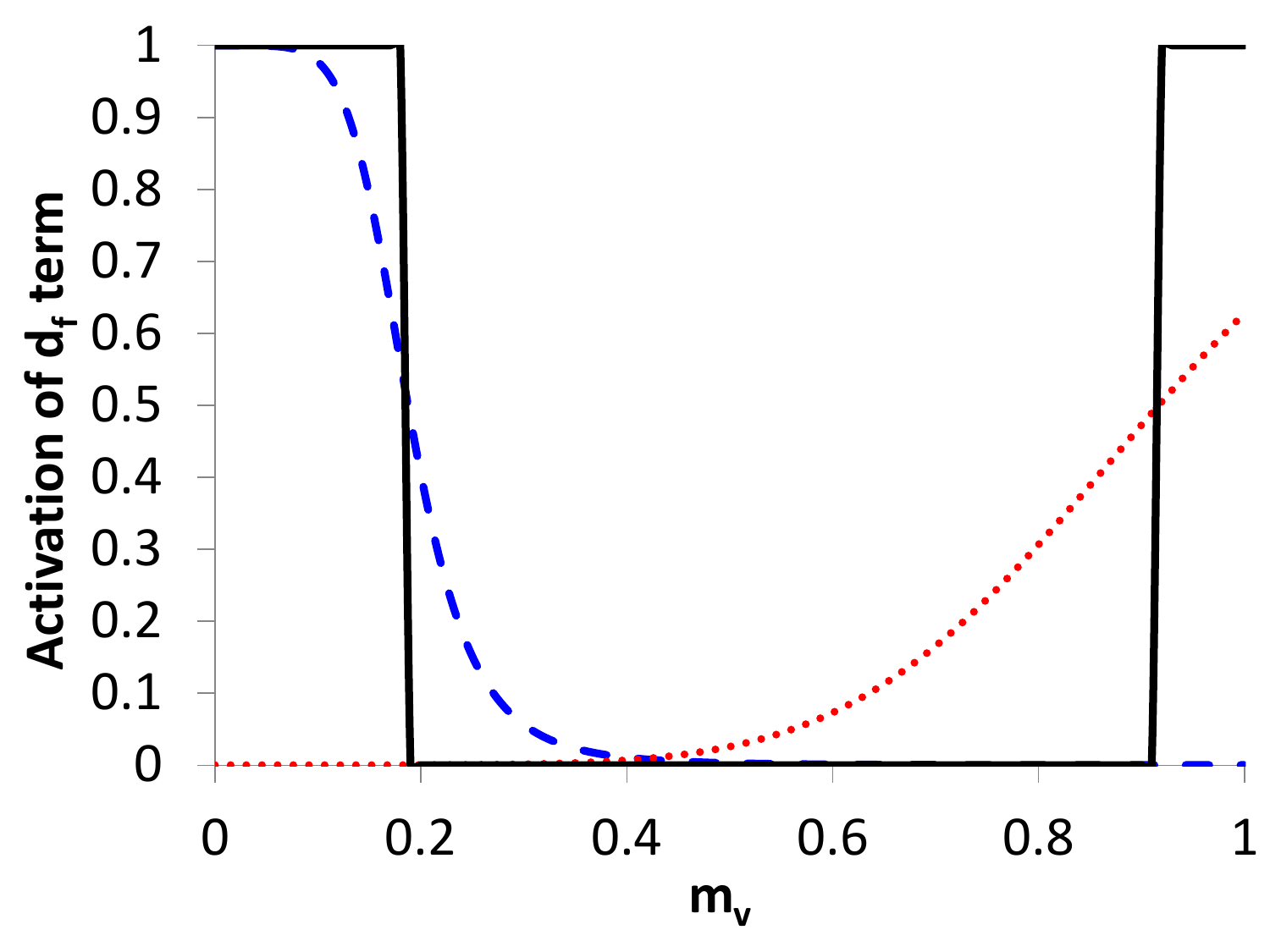}
\end{center}
\caption{Comparison between the activation functions for H-D fibroblast apoptosis using a 6$^{th}$ order Hill function ({\color{blue} $\boldsymbol{---}$}), bFGF-dependent fibroblast apoptosis using a 6$^{th}$ order Hill function ({\color{red} $\boldsymbol{\cdots}$}) and both thresholds modeled with a Heaviside constant function ($\boldsymbol{-}$)}
\label{fig:piecewise}
\end{figure}

Where $\tilde H_{f,hyp}$ is the threshold for H-D apoptosis, $\tilde H_{bFGF}$ is the threshold for the bFGF-dependent apoptosis, $\tilde d_{f,hyp}$ and $\tilde d_{f,bFGF}$ are the corresponding apoptosis rates. This kind of functions are appropriate for modeling thresholds when the variables analyzed are small (as in the reviewed publications). However, when the threshold value is big, the Hill function becomes wide. As can be seen in Figure \ref{fig:piecewise}, the activation function for the bFGF-dependent apoptosis (red dotted line) is much wider than the activation function for the H-D apoptosis (blue dashed line). If Hill functions were used, fibroblasts would undergo apoptosis for almost any condition of vascularization. Similarly, the oxygen-dependent differentiation of osteoblasts from MSCs exhibits a too smooth behavior. As a consequence of this analysis, all thresholds in the present work are modeled with Heaviside functions (black continuous line), which a the same time reduces computational cost. Consequently, equation \ref{eq:threshold} is replaced by:
\begin{small}
\begin{equation}
\tilde d_f  =  \left\{ {\begin{array}{*{20}c}
   {\tilde d_{f,hyp} } & {\begin{array}{*{20}c}
   {} & {if}  \\
\end{array}} & {0 \le \tilde m_v  < \tilde H_{hyp} }  \\
   0 & {\begin{array}{*{20}c}
   {} & {if}  \\
\end{array}} & {\tilde H_{hyp}  \le \tilde m_v  < \tilde H_{bFGF} }  \\
   {\tilde d_{f,bFGF} } & {\begin{array}{*{20}c}
   {} & {if}  \\
\end{array}} & {\tilde H_{bFGF}  \le \tilde m_v }  \\
\end{array}} \right.
\end{equation}
\end{small}

As can be inferred, the parameters in equations \ref{eq:cm}-\ref{eq:gv} can be constant or dependent on the modeled variables. Table \ref{tab:behavior} provides a summary of the non-constant parameters and illustrates their functional behavior. It is important to point out that the bell-shaped curves have different mathematical descriptions depending on the process being modeled (for more information about the mathematical background see \cite{geris2008angiogenesis,olsen1997mathematical,bailon2001mathematical}). The numerical parameters are enunciated in \ref{appendixA}.

\begin{table*}[t] 
  \centering
  \caption{Behavior of non-constant parameters of equations \ref{eq:cm}-\ref{eq:gv} depending on simulated variables.}
  \begin{footnotesize} 
    \begin{tabular}{clll}
    \hline
    \multicolumn{1}{c}{Behavior} & \multicolumn{1}{c}{Phenomenon} & \multicolumn{1}{c}{Y} & \multicolumn{1}{c}{X} \\
    \hline
    Bell-shaped & 	&	&	\\
    {\multirow{4}[1]{*}{\includegraphics[width=0.15\linewidth]{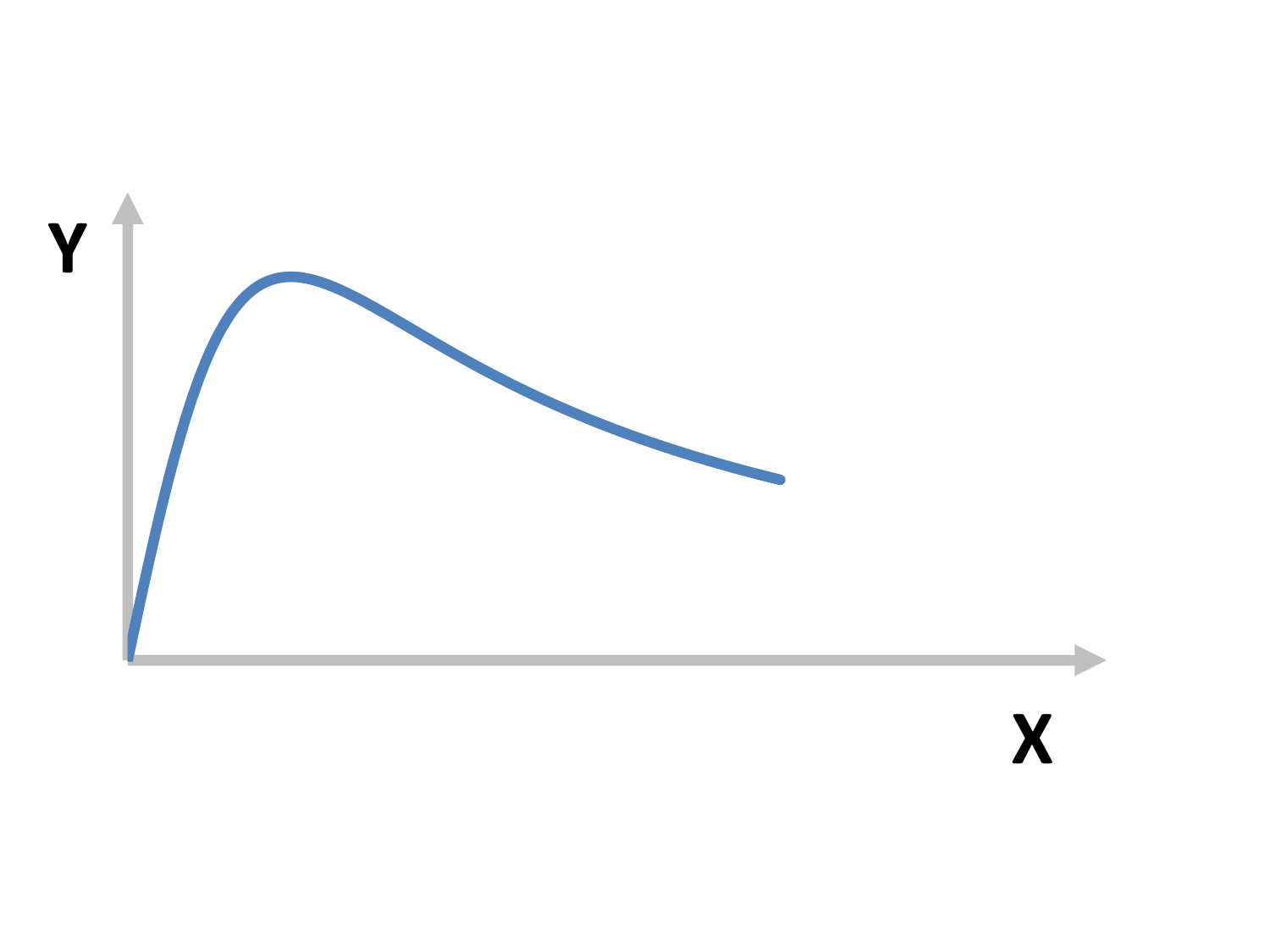}}}
    			& {Haptokinetic diffusion} & {$D_i$ {\footnotesize $(i=m,v)$}} & {$m^*$} \\
       			& {Chemotaxis} & {$C_{i,CT}$ {\footnotesize $(i=m,f,v)$}} & {$g_k$ {\footnotesize $(k=b+v,b,v)$}}  \\
          		& {Haptotaxis} & {$C_{m,HT}$} & {$m^*$} \\
          		& {Cellular proliferation} & {$A_i$} & {$m^*$} \\
          						& 	&	&	\\
    \hline
    Delay-Heaviside 	& 	&	&	\\
    {\multirow{3}[1]{*}{\includegraphics[width=0.15\linewidth]{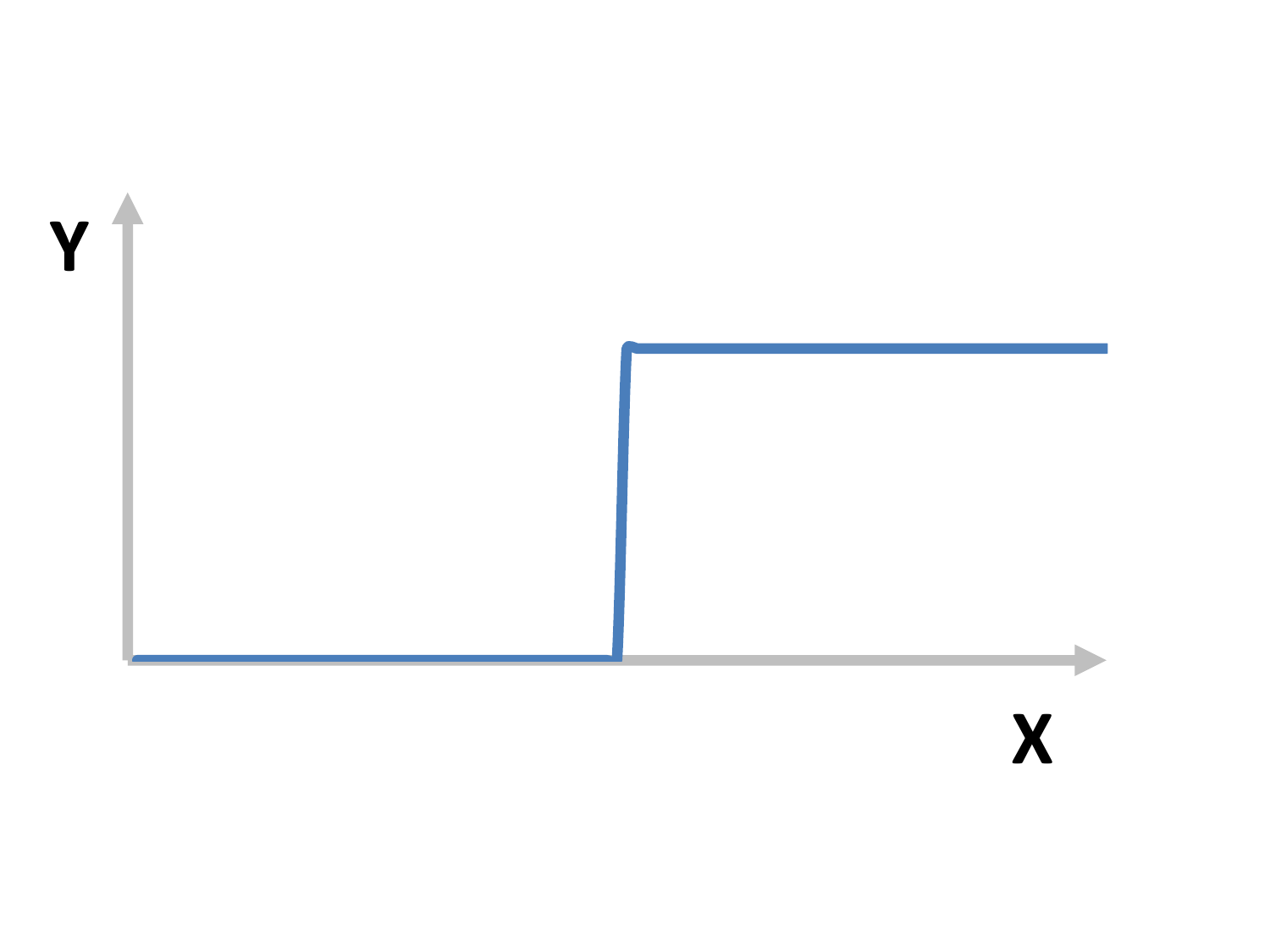}}}     		
				& 	&	&	\\
    			& {bFGF-dependent apoptosis} & {$d_f$} & {$m_v$} \\
           		& {Osteoblast differentiation} & {$F_{mb}$} & {$m_v$} \\
	        	& 	&	&	\\
	        	& 	&	&	\\	        	
    \hline
    1-Delay-Heaviside	& 	&	&	\\
    {\multirow{4}[1]{*}{\includegraphics[width=0.15\linewidth]{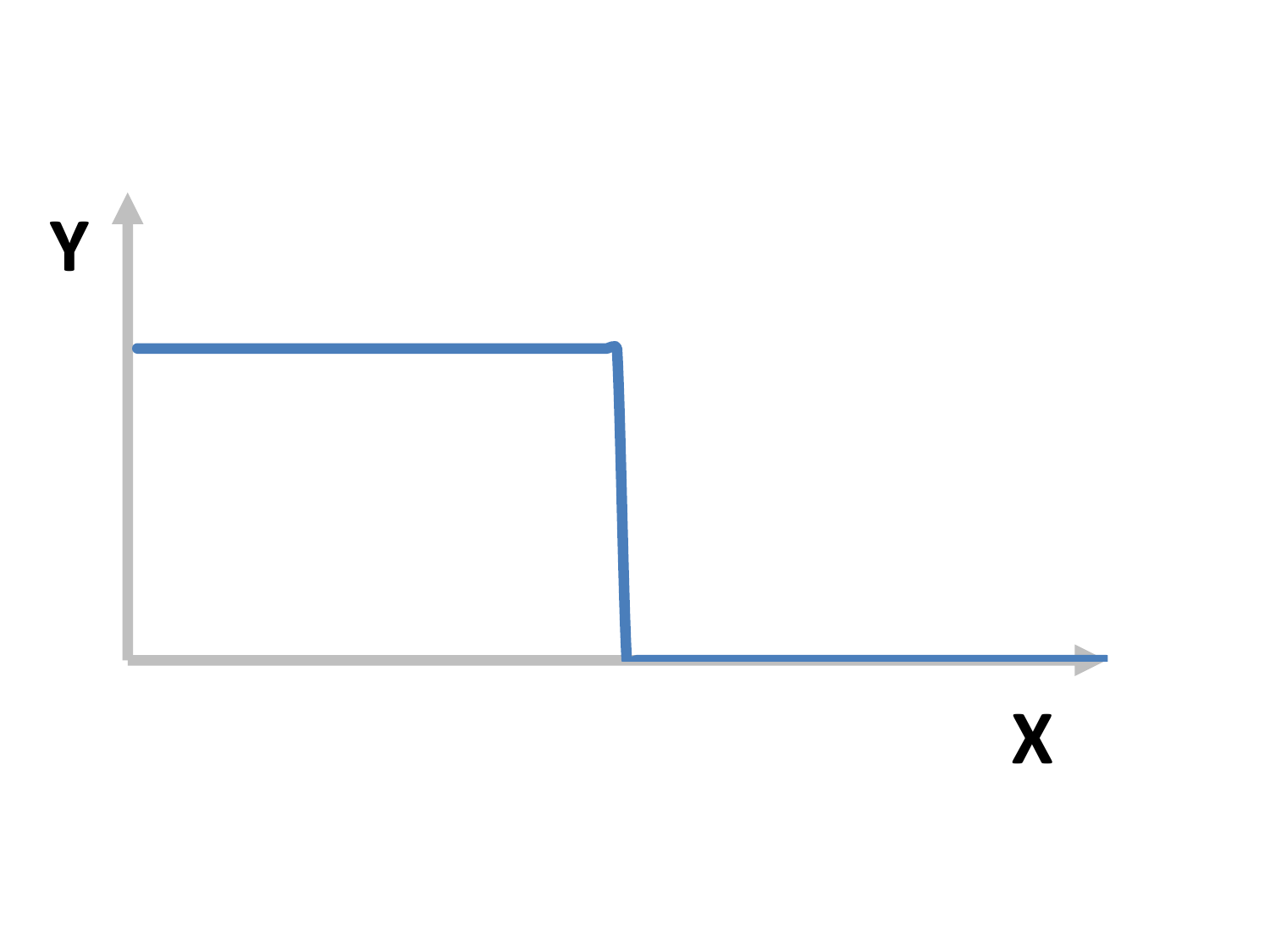}}}

    			& {H-D $g_v$ production} & {$E_{gvi}$} & {$m_v$} \\
       	      	& {H-D apoptosis} & {$d_i$ {\footnotesize $(i=m,f,b)$}} & {$m_v$} \\
          		& {$g_v$ production} by osteoblasts & {$E_{gvb}$} & {$g_v$} \\
	        	& {$g_v$ production}  by BC & {$E_{clot}$} & {$m_v$} \\
	        	& 	&	&	\\
    \hline
    \multicolumn{4}{l}{{\footnotesize $^*$ $m=m_f+m_b$}}\\
    \end{tabular}
   \end{footnotesize}
  \label{tab:behavior}
\end{table*}

\subsubsection{Clot-induced angiogenesis}

Despite of the fact that BC was not considered as an individual matrix, the effect of this temporal tissue on recruiting newly formed vasculature is of outmost importance, and hence included in the model. To this end, a thresholding model was applied for the production of $\tilde g_v$ by BC (see second term in eq. \ref{eq:gv}). This production rate was deemed constant for low values of $m_v$ and zero for high values of $m_v$. 

\subsubsection{Additional considerations}
As in \cite{peiffer2011hybrid,carlier2012mosaic,carlier2015oxygen}, $\tilde g_v$ production by osteoblasts was modeled to be high for low angiogenic growth factor concentrations. These recent studies excluded the original requirement of $g_v$ for the osteoblast differentiation \cite{geris2008angiogenesis}. However, this term was maintained in the present work because of its observed influence in the simulation results and its reported importance \cite{street2002vascular,mayer2005vascular}. Thus, four factors are required for osteoblast differentiation: $c_m$, $m_v$, $g_v$ and $g_b$.  \\

Finally, the differentiation parameters used by Carlier \textit{et al.} \cite{carlier2015oxygen} were adopted. These parameters are listed in Table \ref{tab:diffparam}. Remark that all the thresholds were adapted to the new description of vascularization-dependent effects.\\

Unless it is explicitly indicated in the text, the parameters for equations \ref{eq:cells}-\ref{eq:gf} were defined exactly as by Geris \textit{et al.} \cite{geris2008angiogenesis}. In this sense, the haptokinetic behavior of the diffusivity parameters for cells ($D_i$ in equation \ref{eq:cells}) is modeled as a bell-shaped curve with the following expression:

\begin{small}
\begin{equation}
D_i  = \frac{{D_{hi} m}}{{K_{hi}^2  + m^2 }}
\end{equation}
\end{small}

Following the same methodology as Geris \textit{et al.} \cite{geris2008angiogenesis}, the parameter $D_{hv}$ was adjusted, so that the maximum value of the bell-shaped function corresponds to the diffusivity value reported by Rupnick \cite{rupnick1988quantitative} in their \textit{in-vitro} experiments. Due to the lack of exact quantitative information in the literature, the value of $K_{hv}$ reported by Geris was adopted. The result of the described methodology yields:

\begin{small}
\begin{equation}
D_{hv}  = {\rm{1}}{\rm{.41696}} \times 10^{{\rm{ - 05}}} \frac{g \times mm^2}{{day}}
\end{equation}
\end{small} 

Similarly, the fibroblast diffusivity was taken from the numerical study by Olsen \textit{et al.} \cite{olsen1995mechanochemical}. Remark that all parameters were adimensionalized as described in \ref{appendixA}.

\subsection{Tooth extraction model} \label{expmodel}
The healing of an extraction socket following tooth removal in dogs \cite{cardaropoli2003dynamics} was simulated. In this experimental model, the distal root of the fourth premolar of nine mongrel dogs (12 months old and of about 10 kg) was extracted, damaging the bone tissue structure and triggering a healing process. The dogs were sacrificed at several time points after the extraction and the volume occupied by different types of tissues (BC, GT, PCT, woven bone and mineralized bone) was determined by histomorphometric examination.\\

\begin{figure}[t]
\centering
\resizebox{\linewidth}{!}{
\subfigure[Histological examination at first day after tooth extraction \cite{cardaropoli2003dynamics}.]{\label{histo1} \includegraphics[height=7cm]{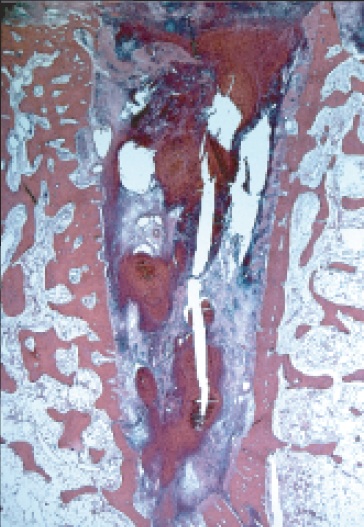}}\qquad
\subfigure[Modeled geometry and mesh. Injured border for boundary conditions shown in red.]{\label{geomsec} \includegraphics[height=7cm]{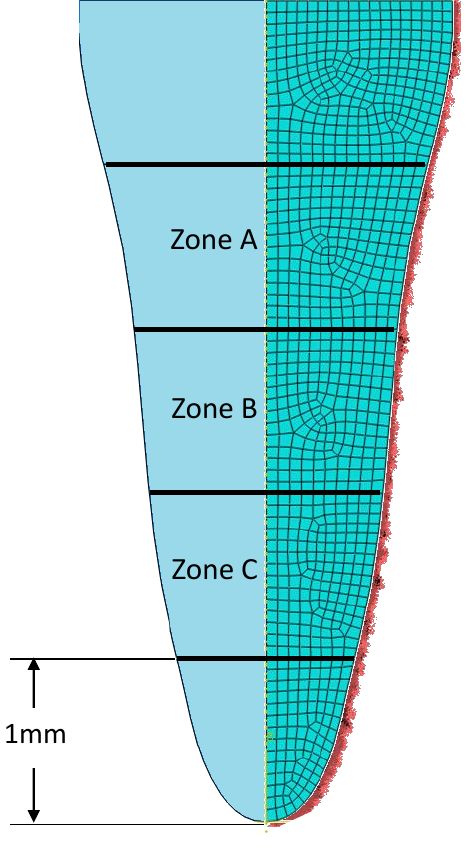}}\ 
}
\caption{Geometry for the simulations.}
\label{fig:Geom}
\end{figure}

The geometry used for the simulations was obtained from a histological examination of the extraction zone taken one day after tooth extraction (Figure \ref{histo1}). The dimensions of the socket and the distribution of the histological sections were derived from the description made by Cardaropoli \textit{et al.} \cite{cardaropoli2003dynamics}, resulting in the representation in Figure \ref{geomsec}, which is 5 mm deep and has a coronal diameter of 2.26 mm. Each histological zone is 1 mm high. These dimensions are coherent with the ones reported by Araújo \textit{et al.} in a dimensional study in mongrel dogs of the same age and size \cite{araujo2005dimensional}.

\subsection{Boundary and initial conditions} \label{boundini}
\begin{figure} 
\centering
\includegraphics[width=0.8\linewidth]{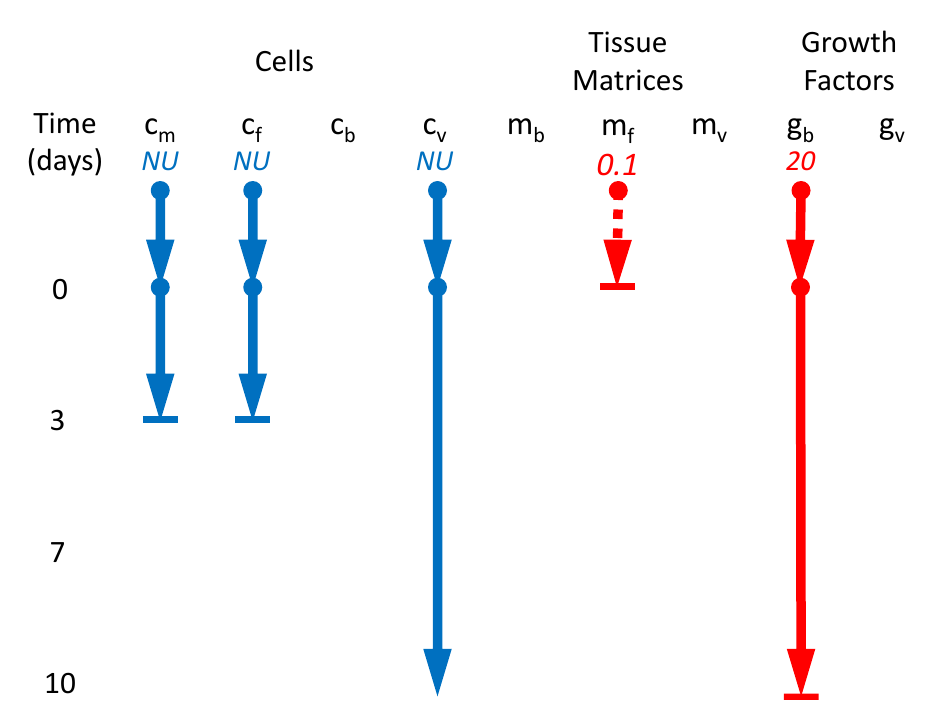}
\caption{Model conditions: graphical representation. 
Continuous and dashed lines represents boundary and domain conditions respectively. Blue lines indicate non-uniform (NU) conditions, which depend on wound depth (Figure \ref{fig:NumCell}), while red lines are defined homogeneously. Arrows ending in a horizontal line mean that the condition is released at that precise moment.
}
\label{fig:conditions}
\end{figure}

Figure \ref{fig:conditions} shows all boundary and initial conditions prescribed. As in previous works \cite{geris2008angiogenesis,peiffer2011hybrid,carlier2012mosaic,carlier2015oxygen}, no-flux boundary conditions (von Neumann) were prescribed on all borders, while Dirichlet boundary conditions were imposed for certain variables on the injured border (Figure \ref{geomsec}) for stipulated timespans. Cardaropoli \textit{et al.} found large numbers of MSCs, fibers and dilated vascular units on the severed PDL\cite{cardaropoli2003dynamics}. Therefore, $\tilde c_m$, $\tilde c_f$ and $\tilde c_v$ were supposed to originate from the surrounding borders. The boundary condition for osteoblasts used by Geris \textit{et al.} \cite{geris2008angiogenesis} was not set (as in \cite{peiffer2011hybrid} onward).\\

McCulloch \textit{et al.} \cite{mcculloch1987paravascular,mcculloch1983cell} studied the communication of endosteal bone with the PDL in mouse alveoli and indicated its direct influence on cell populations. They showed that the percentage of vascular channels (PVC) which communicate with bone marrow varies depending on the depth, being the inner bone richer in vascularization than the outer bone. A regression analysis yielded that the experimental data reported by McCulloch can be approximated using an exponential regression with a high correlation coefficient ($R^2=0.995$). Therefore, we have prescribed the density of MSCs, fibroblasts and endothelial cells in the border adopting the same exponential distribution. The corresponding apical cell densities on the boundary were set at $\tilde c_{m,bc} = \tilde c_{f,bc} = 3.76\times 10^{-3}$ and $\tilde c_{v,bc} =0.188$. These values were calculated taking into account the following:
\begin{itemize}
\item The PVC ratio between alveolar cortex and femoral neck reported by Kingsmill \textit{et al.} \cite{kingslanding} and Young \textit{et al.} \cite{young2008microcomputed}.
\item The boundary conditions established by Geris \textit{et al.} \cite{geris2008angiogenesis} and 
Olsen \textit{et al.} \cite{olsen1997mathematical}.
\end{itemize}

\begin{figure} [h]
\centering
\includegraphics[angle=-90,origin=c, width=0.55\linewidth]{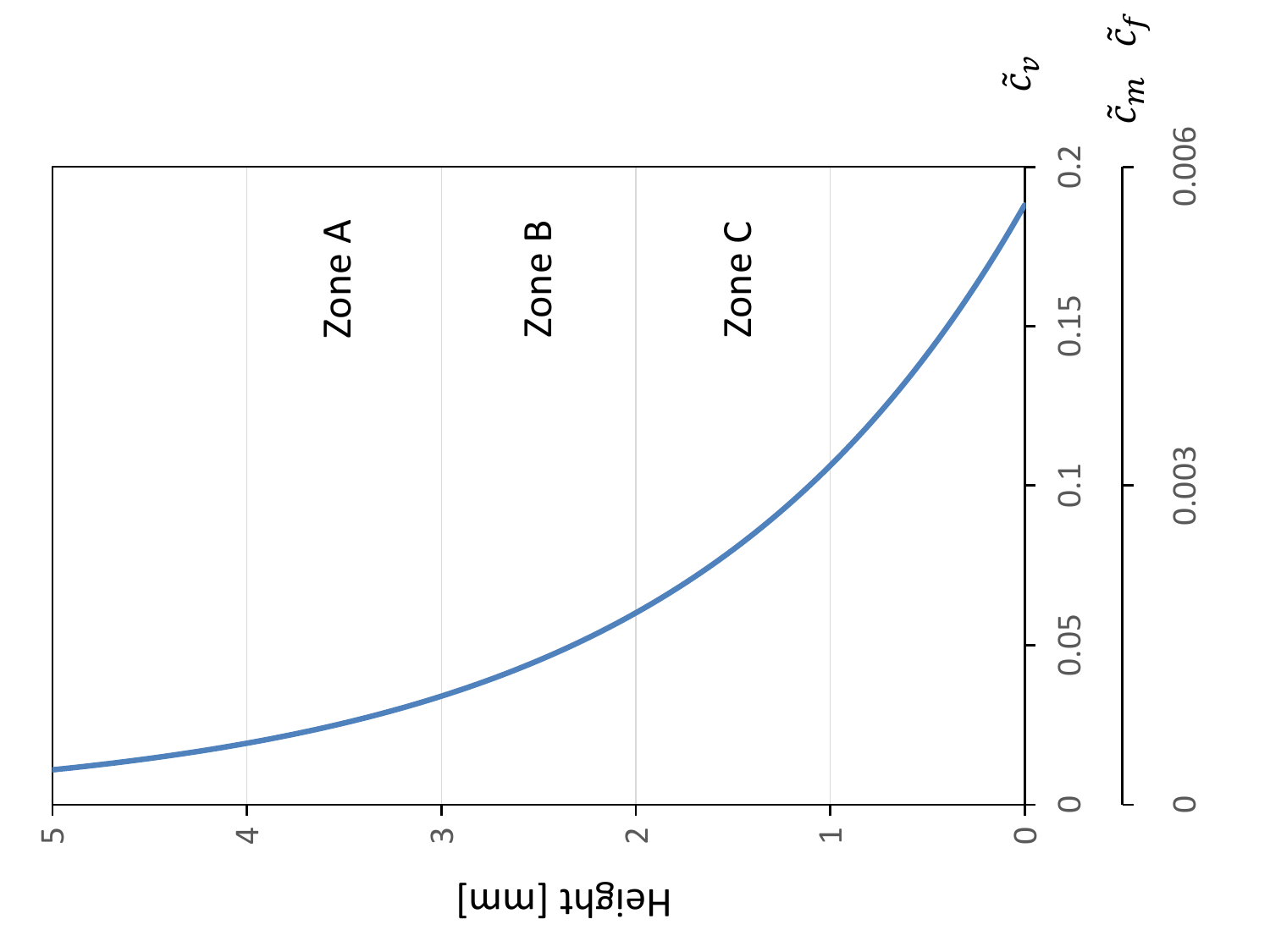}
\caption{Boundary value of MSC, fibroblast and endothelial cell density ($\tilde c_{m,bc}$, $\tilde c_{f,bc}$ and $\tilde c_{v,bc}$ respectively) versus depth of injury. The cell distribution was adjusted using the same functional form obtained  by regression from the experimental data reported by McCulloch \textit{et al.} \cite{mcculloch1983cell,mcculloch1987paravascular} and taking into account the boundary conditions proposed by Geris \textit{et al.} \cite{geris2008angiogenesis} and the PVC ratio from Kingsmill \cite{kingslanding}.}
\label{fig:NumCell}
\end{figure}

Moreover, the calculated PVC ratio between the alveolar most coronal and most apical sections is $5.77\times 10^{-2}$ \cite{mcculloch1987paravascular}. Thereupon, the value for MSCs and fibroblasts on the coronal region was set at $\tilde c_{m,bc} = \tilde c_{f,bc} = 2.1687\times 10^{-4}$ and endothelial cells at $\tilde c_{v,bc} =0.0108$. Figure \ref{fig:NumCell} shows the non-dimensional density of cells prescribed along the injury border. In spite of the presence of endothelial cells on the injured bone end, $\tilde m_v$ was not prescribed as a boundary condition due to the tissue impairment caused during the extraction procedure.\\

Geris \textit{et al.} set $\tilde g_{b,bc}=20$ homogeneously on the injured cortical bone border during the first 10 days \cite{geris2008angiogenesis}. Since the bundle bone (which is mostly cortical) was exposed during the removal of the premolar, this boundary condition was adopted.\\ 

The initial condition for the fibrous matrix was defined as $\tilde m_{f,0}=0.1$ in the whole domain \cite{geris2008angiogenesis} (representing only BC at this time point). A sensitivity analysis showed that the initial condition for $\tilde g_v$ is negligible for the case study. 

\subsection{Numerical implementation}\label{NumImp}
The implementation of the reaction-diffusion-taxis coupled PDE system was achieved by means of FEM. Equations \ref{eq:cm}-\ref{eq:gv} were discretized using the Petrov-Galerkin approach and the code for assembling the system matrices (mass, stiffness and flux)  was written in Fortran 90. In order to ensure mass conservation and improve numerical  performance, a generalized flux-corrected transport algorithm (lumping method) proposed by Strehl \textit{et al.} \cite{strehl2013positivity,strehl2010flux} was used.
Since all analyzed variables must be positive, non-negativity was guaranteed by both lumping and active adaptation of time increments.\\

Half of the modeled geometry was simulated taking into account axisymmetrical behavior. The mesh was made up of 907 quadrilateral elements with bi-linear shape functions (Figure \ref{geomsec}).

\section{Results and discussion} \label{results}

\begin{figure*}[t]
\centering
\includegraphics[width=1\linewidth]{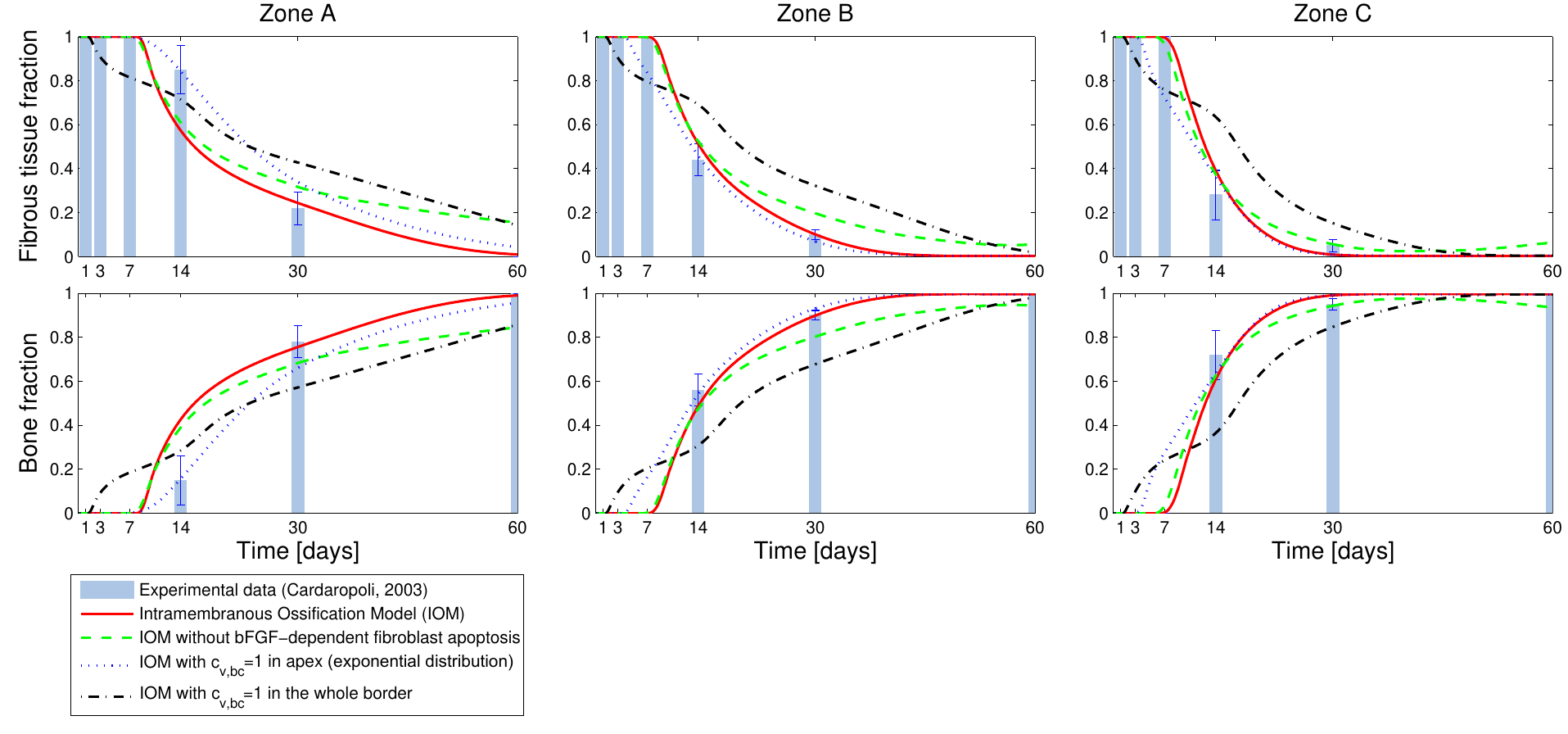}
\caption{Bone and fibrous tissue fractions in alveolus. The bars (average $\pm$ standard deviation) represent the histologically measured values at 1, 3, 7, 14, 30 and 60 days after tooth extraction (see section \ref{expmodel}). The curves are the tissue fractions predicted by the Intramembranous Ossification Model (IOM) and its modifications.}
\label{fig:curves}
\end{figure*}

Bars in Figure \ref{fig:curves} show the histologically measured tissue fractions \cite{cardaropoli2003dynamics} for fibrous tissue and bone in histological zones A, B and C at days 1, 3, 7, 14, 30 and 60 after tooth extraction. The computationally predicted tissue fractions are illustrated as follows: IOM (red continuous line), IOM without bFGF-dependent fibroblast apoptosis (green dashed line) and IOM using different boundary conditions (blue dotted line and black dash-dot line). For the intermediate states (days 14 and 30) the error bars indicate the standard deviation of the experimental data, whereas at days 1, 3 and 7 only fibrous tissue was observed and at day 60 only bone tissue was reported \cite{cardaropoli2003dynamics}.\\

The predicted tissue fractions in each node were obtained with the following mathematical expression:\\
\begin{small}
\begin{equation}
\chi_j  = \frac{{\tilde m_j }}{{\tilde m_f  + \tilde m_b }}
\end{equation}\\
\end{small}
Afterwards, the average tissue fraction for all the nodes located in each histological zone was computed in every time increment, in order to generate the curves that are shown in Figure \ref{fig:curves}. The mean absolute error for the IOM is 3.04\%, for the green dashed line 5.18\%, for the blue dotted line 4.40\% and for the black line 13.1\%. 

\subsection{Intramembranous ossification model (IOM)}
The ossification process starts shortly after day 7, which is consistent with the data of Cardaropoli \textit{et al.} \cite{cardaropoli2003dynamics}. The ossification rate increases suddenly and then slowly decreases (see also Figure \ref{fig:Diff}). Around day 14, angiogenic factors ($\tilde g_v$) are depleted and the osteoblast differentiation stops. Consequently, the population of osteoblasts can only increase by means of proliferation. This and the synthesis saturation expressed in equation \ref{eq:mb} cause a deceleration in bone formation.\\

\begin{figure*}[t]
\begin{center}
	\includegraphics[scale=1]{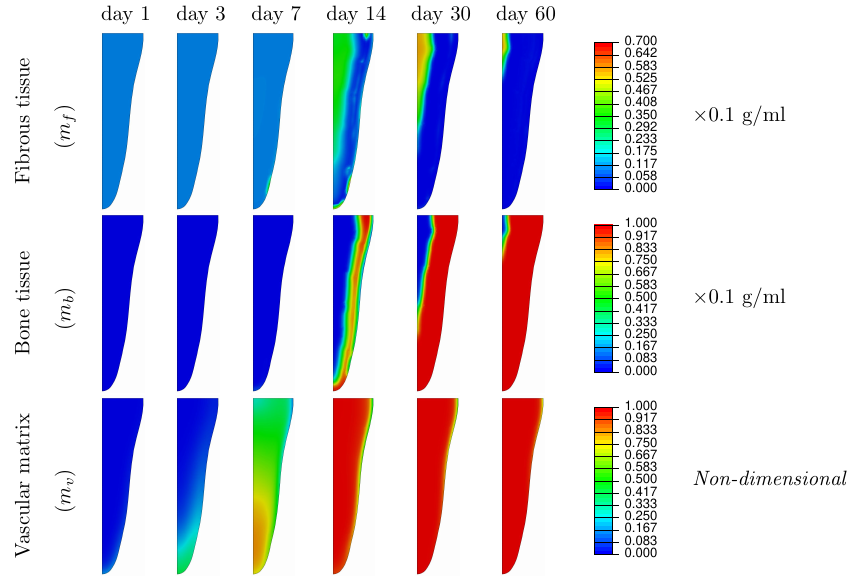}
\end{center}
\caption{Spatiotemporal distribution of ECMs: Fibrous tissue ($m_f$), bone ($m_b$) and vascular tissue ($m_v$).}
\label{fig:spatemp_m}
\end{figure*}

Note that the ossification is triggered first in \textit{Zone C}, then in \textit{Zone B} and lastly in \textit{Zone A} and that the bone formation rate is slower in the upper zones. This is a consequence of the higher concentration of $\tilde c_m$ and $\tilde m_v$ (produced by the prescribed boundary conditions), and  the geometrical characteristics of the different zones. Since the apical regions are smaller, the initial ossification on the surrounding borders (Figure \ref{fig:spatemp_m}) results in an accentuated increase in bone fraction.\\

The spatio-temporal distribution of tissues obtained in the simulations is shown in Figure \ref{fig:spatemp_m} and the contours for cells and growth factors are shown in Figures \ref{fig:spatemp_c} and \ref{fig:spatemp_g}, respectively. It can be seen that fibrous tissue is degraded as bone is formed from the injured borders inwards and bottom up, in accordance with the investigation of Cardaropoli \textit{et al.} \cite{cardaropoli2003dynamics}.

\subsubsection{Days 0-14}
The gradient of cells described by the exponential definition on the boundary induces bottom up and inward diffusion. MSCs and fibroblasts invade the socket area from the injured bone ends secreting angiogenic growth factors due to hypoxia, which along with the $\tilde g_v$ produced by BC fills up the entire domain. Meanwhile, the autocrine and paracrine role of angiogenic growth factors expressed by the MSCs promotes clustering in the center of the socket. Further concentration of MSCs is seen on the wound edge as a result of the chemotactic response to osteogenic growth factors.\\

During the early stages of wound healing, the big amounts of the produced angiogenic growth factors degrade rapidly and are consumed by the endothelial cells during angiogenesis. At day 7, $\tilde g_v$ is almost completely consumed and some clusters of endothelial cells can be observed as a vascularization-forming traveling wave advances through the domain. Remark that the wave comprises two advancing peaks that propagate through the socket, whose maximum values are slightly greater than $\tilde c_v=1$ (saturation value for proliferation) in zones where chemotactic and haptotactic effects are significant.\\

As soon as the boundary condition for $\tilde g_b$ is released, the border MSCs begin to migrate triggering the ossification front advance. Meanwhile, the central cluster mainly differentiates into fibroblasts favored by the absence of $g_b$. At the 14$^{th}$ day,  fibrous tissue synthesized by the cluster of fibroblasts is observed in the wound center as well as in two isolated points near the border of the apical and coronal zones. This latter formation can be explained by both the differentiation of MSCs into fibroblasts and a previous chemotactic migration of $c_f$ towards high $\tilde g_b$ spots, which are favored by the absence of fibroblast apoptosis (the vascularization is neither too high nor too low to induce it).

\subsubsection{Days 14-30}
\begin{figure*}[t] 

\begin{center}
	\includegraphics[scale=1]{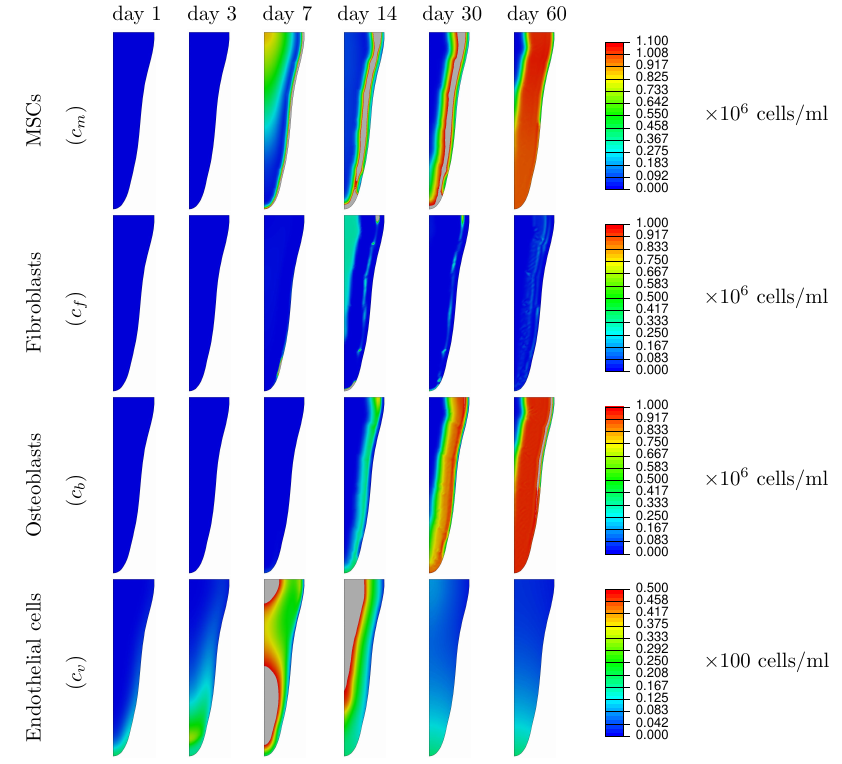}
\end{center}
\caption{Spatiotemporal distribution of cells: MSCs ($c_m$), fibroblasts ($c_f$), osteoblasts ($c_b$) and endothelial cells ($c_v$).}
\label{fig:spatemp_c}
\end{figure*}

\begin{figure*} 

\begin{center}
	\includegraphics[scale=1]{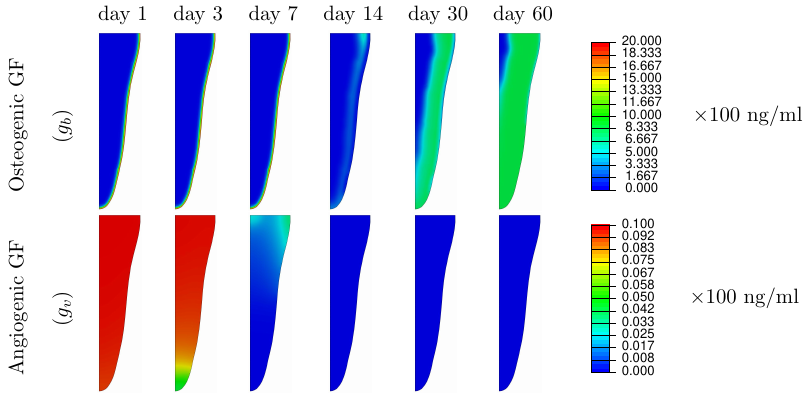}
\end{center}
\caption{Spatiotemporal distribution of growth factors: osteogenic growth factor ($g_b$) and angiogenic growth factor ($g_v$).}
\label{fig:spatemp_g}
\end{figure*}

Between days 14 and 30, vascular ECM is formed from the root canal upwards, filling almost completely the whole extraction socket by day 14$^{th}$. This causes bFGF-dependent fibroblast apoptosis to become dominant, which, together with the presence of osteoblasts, leads to a rapid resorption of fibrous tissue. At this time, the ossification front advances over the domain towards the center of the injury from the apical to the coronal zones. As can be seen in Figures \ref{fig:spatemp_c} and \ref{fig:spatemp_g}, the osteogenic growth factor distribution is similar to the osteoblasts', which illustrates the autocrine and paracrine behavior.\\ 

The decreasing population of fibroblasts clustered in the central coronal area achieves to densify the fibrous matrix in this area. When the formation of blood vessels is concluded, the distribution of endothelial cells accommodates to resemble the boundary condition.

\subsubsection{Days 30-60}

Between days 30 and 60, the ossification front advances continuously as expected. The deceased fibroblasts on the central coronal zone left behind a densified fibrous tissue in that area. At day 60, the bone tissue is not yet formed at the coronal center of the socket, as can be observed in the histologies reported by Cardaropoli \cite{cardaropoli2003dynamics}. At the end of the simulation, MSCs have proliferated through the whole domain populating the newly formed bone marrow, making possible the subsequent processes, such as bone remodeling and eventually wound healing.\\

\begin{figure}[h]
\centering
\includegraphics[width=\linewidth]{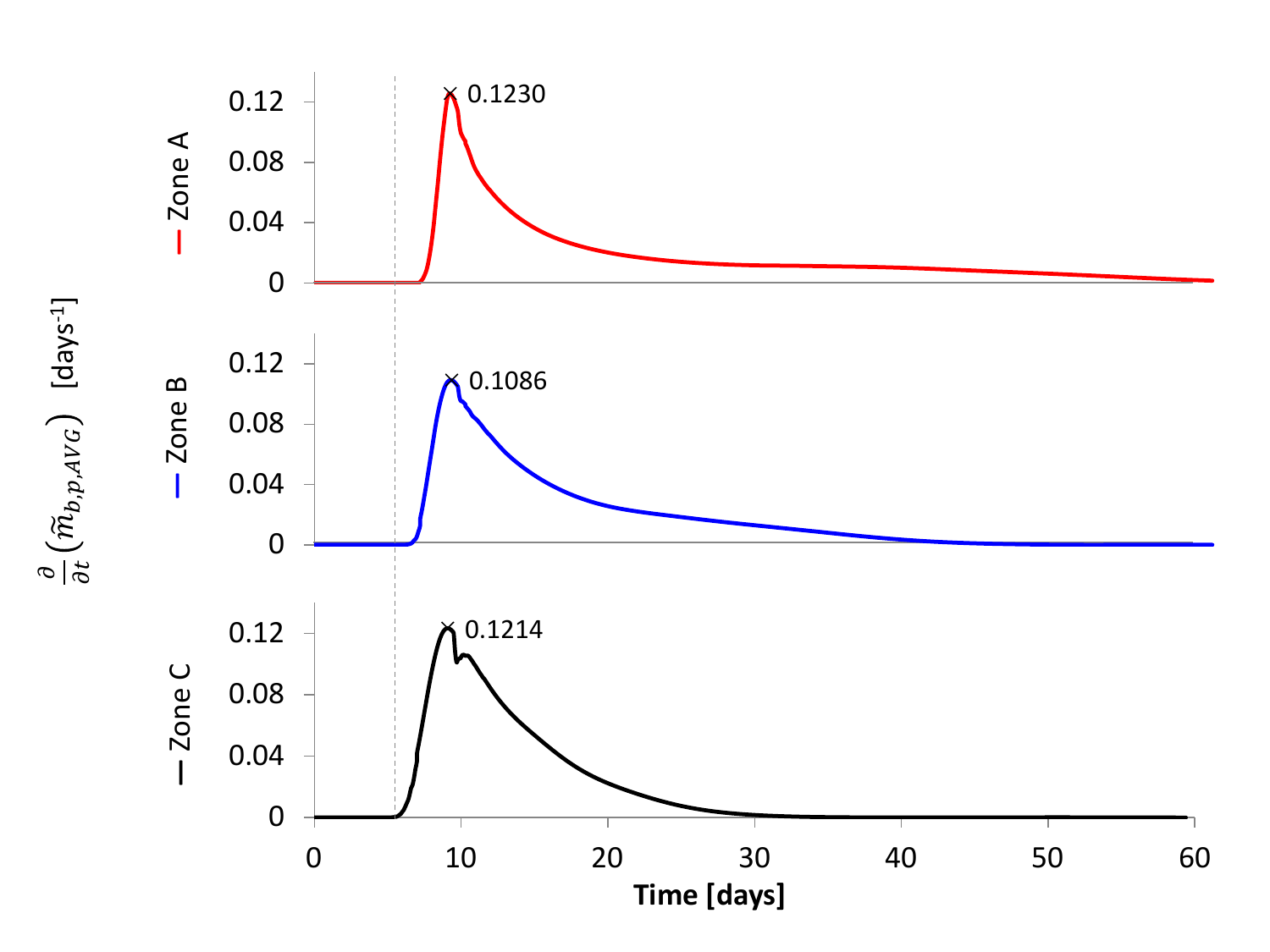}
\caption{Derivatives of the averaged tissue fractions.}
\label{fig:Diff}
\end{figure}

Figure \ref{fig:Diff} illustrates bone formation rates (BFR) for the three alveolar zones. After the ossification process is triggered at around day 7, the inital BFR is similar for all zones. The peak BFRs are 12.30\%/day, 10.86\%/day and 12.14\%/day for \textit{Zone A}, \textit{B} and \textit{C}, respectively and occur around day 10. \\

\subsection{Modifications to the IOM}

The green dashed lined of Figure \ref{fig:curves} depicts that the absence of bFGF-dependent fibroblast apoptosis produces a late and unexpected densification of the fibrous tissue in all zones, leading to an increase in fibrous matrix proportion in comparison with bone tissue. Without the apoptosis term, fibroblasts continue to proliferate and create excessive ECM, as can be seen in \textit{Zone C}, where a growth trend of fibrous tissue is shown.\\

The blue dotted line in Figure \ref{fig:curves} shows the IOM predictions when the boundary condition is changed. The exponential function for cell density on the injured border (section \ref{boundini}) was used, but the boundary values for $c_m$, $c_f$ and $c_v$ were not scaled to account for the difference in vascularization between long bones and the mandible. This approach seems to fit better the histological observations in \textit{Zone A}, but it is detrimental for the prediction in \textit{Zones B} and \textit{C}. Since there are many endothelial cells on the boundary, vascularization is formed faster than in the IOM. In \textit{Zone A}, endothelial cells reach the spots with high concentration of $g_v$ and consume it before there is enough $g_b$ to allow osteoblast differentiation.\\

In \textit{Zones B} and \textit{C}, bone formation and the correspondent fibrous tissue degradation are very similar to the IOM from day 14 onward. However, bone formation begins much earlier. This phenomenon responds to the fact that three of the four required conditions for osteoblast differentiation are magnified: the amount of MSCs on the apical region, the production of angiogenic growth factor and the formation of vascular ECM. Approximately at day 14, endothelial cells regulate the concentration of angiogenic growth factors on the apical region by consuming them. By doing so, osteoblast differentiation slows down, as in the IOM.\\

For the black dash-dot line in Figure \ref{fig:curves}, the boundary condition for $c_m$, $c_f$ and $c_v$  was uniform and unscaled. The early bone formation that was observed in the blue dotted line is more accentuated and also present in \textit{Zone A}. Then, the BFR slows down, producing a pronounced mismatch with the experimental data. The importance of an accurate description of the boundary values becomes clear.

\subsection{Limitations of the IOM} \label{limitations}
As can be observed in Figure \ref{fig:curves}, there is a mismatch with the experimental data at day 14 in \textit{Zone A}. These dissimilitudes may be a consequence of not simulating the stages of clot formation and resorption. Even though the blood clot, the provisional matrix and the granulation tissue are all fibrous tissues, they are formed and degraded at different time points of the bone formation process. Experimental studies \cite{cardaropoli2003dynamics,amler1969time} show that the big primary BC needs to be degraded in order to leave space for PCT and GT, which serve as scaffolds for cell migration and consequent vascularization \cite{chevrier2007chitosan}. Since there would be a greater volume of BC to resorb in \textit{Zone A}, the incorporation of BC dynamics might result in a delayed bone formation and will be the subject of future work. \\

The discrepancy in tissue fractions in \textit{Zone C} at day 30 may be explained by the high osteoblast density over the areas of recently formed bone (Figure \ref{fig:spatemp_c}). In order to avoid this effect, a rigorous review of the mechanisms that underlie the regulation of the osteoblast population via apoptosis and osteocyte differentiation at high woven bone densities ought to be made. Another possible explanation for the disparities is that the trans-differentiation of fibroblasts into osteoblasts during the intramembranous ossification \cite{lin1994differentiation,hee2006influence} was not considered.\\

In the IOM, the saturation terms for cell proliferation (equations \ref{eq:cm}-\ref{eq:cv}) are only dependent on each cell population. However, due to the existent competition for space and nutrients, saturation limits for cell proliferation depend not only on the population of a given cell type but also on that of other kinds of cells. Similarly, the synthesis saturation for each ECM (equations \ref{eq:mf}-\ref{eq:mv}) considers only the specific tissue maximum density. A revision of the saturation mathematical terms will be addressed in future studies.\\

Since the processes that take place in the surrounding tissue were not simulated, their effect on the bone formation inside the alveolus had to be captured with boundary conditions (as in \cite{geris2008angiogenesis,peiffer2011hybrid,carlier2012mosaic,carlier2015oxygen}). However, the surrounding bone marrow and cortical bone are influenced by both mechanical and biological processes. In point of fact, Araújo \textit{et al.} \cite{araujo2005dimensional} illustrates how the alveolar ridge dimensions are altered after the tooth extraction on 10 kg mongrel dogs. In the absence of a tooth, bone resorption caused by low mechanical loading can be observed and a retraction of the neighboring bone marrow to more apical zones may be expected. As seen in section \ref{boundini}, this would modify the injured PDL cell populations, \textit{i.e.}, the boundary conditions should not be constant but time dependent. \\

The resorption and elimination of fibrous tissue are a consequence of both the metalloproteinases produced by osteoblasts and the action of macrophages \cite{meikle1992human}, which enter the domain along with the osteoblasts. However, due to the lack of quantitative data, fibrous matrix degradation was modeled as proportional to the osteoblast concentration (equation \ref{eq:mf}). By doing so, bone healing was supposed to occur as in a healthy individual. For some pathologies, the action of the macrophages may have to be modeled separately. Further research is needed in order to describe in more depth the processes that underlie fibrous tissue degradation.\\

Due to the lack of information about the 3D characteristics of the alveoli, the simulation domain was modeled from a 2D-histology and supposed to have axisymmetric behavior. This could have introduced some error in the outcome. The mathematical model and the computational implementation presented in this article could also be implemented in geometries obtained from 3D medical imaging tools (such as CT-scan or MRI). Nonetheless, the computational cost would be much higher.\\

Fluctuations in oxygen tension due to consumption by cells may occur. Since oxygen tension was not directly simulated these fluctuations cannot be observed in the present study.  A possible manner to overcome this issue is to use an approach like the one shown by Carlier \textit{et al.} \cite{carlier2015oxygen}.\\

Future models must take into consideration the individual action of the different growth factors in order to better depict the biological phenomena. For instance, even though it has been reported that bFGF acts as an angiogenic growth factor (it has autocrine and paracrine effects on endothelial cells \cite{schweigerer1987capillary,vlodavsky1987endothelial}), the new fibroblast apoptosis  term was not implemented as dependent on $\tilde g_v$. The reason is that fibroblast apoptosis would be incorrectly induced in early stages of the bone healing process, since all the cells produce VEGF (another angiogenic growth factor) in hypoxia conditions. Ergo, a more precise model would differenciate between VEGF and bFGF. Similar behaviors can be identified for the BMPs (1–8), GDFs (1, 5, 8, and 10), and TGF$\beta$1, which contribute to both endochondral and intramembranous ossification \cite{gerstenfeld2003fracture}.\\

In the present approach, as in many of the cited works, mesenchymal stem cell differentiation into other types of cells is modeled as a reaction term. This assumes that the differentiation takes place immediately. However, cell differentiation is a gradual process, in which cells pass through a series of intermediate stages by expressing different phenotypes under the influence of both mechanical and biochemical signals\cite{long2012building}. A way of dealing with this phenomenon could be keeping track of the gradual change of cell phenotype (as in the work by Prokharau \textit{et al.} \cite{prokharau2014mathematical}). Another solution to the problem could be to explicitly include some of the intermediate stages of differentiation in the model. These approaches will be addressed in future studies. 

\section{Conclusions}

\begin{itemize}
\item A new mathematical model for intramembranous ossification was developed and validated with experimental data taken from the literature \cite{cardaropoli2003dynamics}. It is, to the best of our knowledge, the first model that describes intramembranous ossification with emphasis on the processes that only occur in this type of ossification.

\item The \textit{in-silico} study of bone healing in mouth can contribute not only to a better understanding of intramembranous ossification but also to the better understanding of the ossification processes in long bones, since some biological phenomena are excluded and the remaining can be analyzed more deeply. 

\item The proposed model and its future improvements will aid planning dental procedures, analyzing new treatment strategies and designing enhanced dental implants. Furthermore, they may not only contribute to reduce the need for animal tests, but also to the proper experimental design for both \textit{in-vivo} and \textit{in-vitro} tests.

\item The simulations showed the importance of considering both H-D and bFGF-dependent fibroblast apotosis, especially for modeling intramembranous ossification. 

\item  Hill funtions are useful for describing activation thresholds, but in furtherance of solving them inside a FEM it is important to be careful with their usage in behalf of numerical concordance.

\item In the direction of improving the IOM, further experimental research is needed in order to accurately measure diffusion coefficients, cell half-life times, growth factor synthesis, consumption and degradation rates, etc. The accuracy of the simulations highly depends on using reliable data.

\item In pursuance of a better understanding of the biological process of bone healing, a rigorous study about other important phenomena should be made: blood clot formation and resorption, specific growth factor synthesis and degradation, macrophage activity, and mechanically induced osteogenesis.

\item As seen in Section \ref{results}, the assigned boundary conditions have an important effect on the simulation outcome. In order to extend the application of this model to the clinical practice, medical imaging techniques jointly with biomarkers assays should be developed to determine the particular characteristics  of each individual and thus achieving customized, more effective therapeutical approaches.

\item Mechanical loads and interstitial fluid flow have an effect on cellular differentiation, migration, proliferation and protein synthesis \cite{gefen2011cellular,thompson2002model,prendergast2010computational}. The omission of these effects in the IOM might have a limited impact on the present analysis, since in the original study Cardaropoli \textit{et al.} \cite{cardaropoli2003dynamics} ensured minimal loading conditions by means of feeding the dogs with a soft pellet diet throughout the experiment. Nonetheless, indirect loading from surrounding teeth and stresses due to possible inflammatory responses will take place in the extraction site. More investigation on the mechanobiological pathways of bone metabolism is needed in order to incorporate these phenomena into the IOM.\\ 

Mechanical loading will play an important role in future \textit{in-silico} studies. This could lead to accurate descriptions of compromised healing situations, of the effects of dental implant loading on bone formation, and of inflammatory responses and their influence on bone healing (including the inflammatory responses caused by growth factors administration \cite{ribeiro2015silico}). \\

\end{itemize}

\section{Conflicts of interest}
The authors declare that they have no conflicts of interest.

\section{Acknowledgements}
The research group would like to thank professor Liesbet Geris for her advice on topics related with the bio-regulatory model and for her friendly disposition and willingness to help.\\

We would also like to thank the Colombian Administrative Department of Science, Technology and Innovation (COLCIENCIAS) for the financial support given to the project.\\

\section*{References}

\bibliography{mybibfile}

\begin{thebibliography}{10}
\expandafter\ifx\csname url\endcsname\relax
  \def\url#1{\texttt{#1}}\fi
\expandafter\ifx\csname urlprefix\endcsname\relax\def\urlprefix{URL }\fi
\expandafter\ifx\csname href\endcsname\relax
  \def\href#1#2{#2} \def\path#1{#1}\fi

\bibitem{luginbuehl2004localized}
V.~Luginbuehl, L.~Meinel, H.~P. Merkle, B.~Gander, Localized delivery of growth factors for bone repair, European Journal of Pharmaceutics and Biopharmaceutics 58~(2) (2004) 197--208.
\newblock \href {https://doi.org/10.1016/j.ejpb.2004.03.004} {\path{doi:10.1016/j.ejpb.2004.03.004}}.

\bibitem{mitragotri2005healing}
S.~Mitragotri, Healing sound: the use of ultrasound in drug delivery and other therapeutic applications, Nature Reviews Drug Discovery 4~(3) (2005) 255--260.
\newblock \href {https://doi.org/10.1038/nrd1662} {\path{doi:10.1038/nrd1662}}.

\bibitem{doblare2004modelling}
M.~Doblar{\'e}, J.~Garc{\i}a, M.~G{\'o}mez, Modelling bone tissue fracture and healing: a review, Engineering Fracture Mechanics 71~(13) (2004) 1809--1840.
\newblock \href {https://doi.org/10.1016/j.engfracmech.2003.08.003} {\path{doi:10.1016/j.engfracmech.2003.08.003}}.

\bibitem{isaksson2012recent}
H.~Isaksson, Recent advances in mechanobiological modeling of bone regeneration, Mechanics Research Communications 42 (2012) 22--31.
\newblock \href {https://doi.org/10.1016/j.mechrescom.2011.11.006} {\path{doi:10.1016/j.mechrescom.2011.11.006}}.

\bibitem{lacroix2002mechano}
D.~Lacroix, P.~Prendergast, A mechano-regulation model for tissue differentiation during fracture healing: analysis of gap size and loading, Journal of biomechanics 35~(9) (2002) 1163--1171.
\newblock \href {https://doi.org/10.1016/S0021-9290(02)00086-6} {\path{doi:10.1016/S0021-9290(02)00086-6}}.

\bibitem{gomez2005influence}
M.~Gomez-Benito, J.~Garcia-Aznar, J.~Kuiper, M.~Doblar{\'e}, Influence of fracture gap size on the pattern of long bone healing: a computational study, Journal of theoretical biology 235~(1) (2005) 105--119.
\newblock \href {https://doi.org/10.1016/j.jtbi.2004.12.023} {\path{doi:10.1016/j.jtbi.2004.12.023}}.

\bibitem{bailon2001mathematical}
A.~Bailon-Plaza, M.~C. Van Der~Meulen, A mathematical framework to study the effects of growth factor influences on fracture healing, Journal of Theoretical Biology 212~(2) (2001) 191--209.
\newblock \href {https://doi.org/10.1006/jtbi.2001.2372} {\path{doi:10.1006/jtbi.2001.2372}}.

\bibitem{geris2010connecting}
L.~Geris, J.~Vander~Sloten, H.~Van~Oosterwyck, Connecting biology and mechanics in fracture healing: an integrated mathematical modeling framework for the study of nonunions, Biomechanics and modeling in mechanobiology 9~(6) (2010) 713--724.
\newblock \href {https://doi.org/10.1007/s10237-010-0208-8} {\path{doi:10.1007/s10237-010-0208-8}}.

\bibitem{garcia2007computational}
J.~Garcia-Aznar, J.~Kuiper, M.~G{\'o}mez-Benito, M.~Doblar{\'e}, J.~Richardson, Computational simulation of fracture healing: influence of interfragmentary movement on the callus growth, Journal of biomechanics 40~(7) (2007) 1467--1476.
\newblock \href {https://doi.org/10.1016/j.jbiomech.2006.06.013} {\path{doi:10.1016/j.jbiomech.2006.06.013}}.

\bibitem{pivonka2010mathematical}
P.~Pivonka, S.~V. Komarova, Mathematical modeling in bone biology: From intracellular signaling to tissue mechanics, Bone 47~(2) (2010) 181--189.
\newblock \href {https://doi.org/10.1016/j.bone.2010.04.601} {\path{doi:10.1016/j.bone.2010.04.601}}.

\bibitem{komarova2003mathematical}
S.~V. Komarova, R.~J. Smith, S.~J. Dixon, S.~M. Sims, L.~M. Wahl, Mathematical model predicts a critical role for osteoclast autocrine regulation in the control of bone remodeling, Bone 33~(2) (2003) 206--215.
\newblock \href {https://doi.org/10.1016/S8756-3282(03)00157-1} {\path{doi:10.1016/S8756-3282(03)00157-1}}.

\bibitem{ribeiro2015silico}
F.~O. Ribeiro, M.~J. G{\'o}mez-Benito, J.~Folgado, P.~R. Fernandes, J.~M. Garc{\'\i}a-Aznar, In silico mechano-chemical model of bone healing for the regeneration of critical defects: the effect of bmp-2, PloS one 10~(6) (2015) e0127722.
\newblock \href {https://doi.org/10.1371/journal.pone.0127722} {\path{doi:10.1371/journal.pone.0127722}}.

\bibitem{geris2008angiogenesis}
L.~Geris, A.~Gerisch, J.~Vander~Sloten, R.~Weiner, H.~Van~Oosterwyck, Angiogenesis in bone fracture healing: a bioregulatory model, Journal of theoretical biology 251~(1) (2008) 137--158.
\newblock \href {https://doi.org/10.1016/j.jtbi.2007.11.008} {\path{doi:10.1016/j.jtbi.2007.11.008}}.

\bibitem{peiffer2011hybrid}
V.~Peiffer, A.~Gerisch, D.~Vandepitte, H.~Van~Oosterwyck, L.~Geris, A hybrid bioregulatory model of angiogenesis during bone fracture healing, Biomechanics and modeling in mechanobiology 10~(3) (2011) 383--395.
\newblock \href {https://doi.org/10.1007/s10237-010-0241-7} {\path{doi:10.1007/s10237-010-0241-7}}.

\bibitem{carlier2012mosaic}
A.~Carlier, L.~Geris, K.~Bentley, G.~Carmeliet, P.~Carmeliet, H.~Van~Oosterwyck, Mosaic: a multiscale model of osteogenesis and sprouting angiogenesis with lateral inhibition of endothelial cells, PLOS: Computational Biology (2012).
\newblock \href {https://doi.org/10.1371/journal.pcbi.1002724} {\path{doi:10.1371/journal.pcbi.1002724}}.

\bibitem{zhang2012role}
L.~Zhang, M.~Richardson, P.~Mendis, Role of chemical and mechanical stimuli in mediating bone fracture healing, Clinical and Experimental Pharmacology and Physiology 39~(8) (2012) 706--710.
\newblock \href {https://doi.org/10.1111/j.1440-1681.2011.05652.x} {\path{doi:10.1111/j.1440-1681.2011.05652.x}}.

\bibitem{gerstenfeld2003fracture}
L.~C. Gerstenfeld, D.~M. Cullinane, G.~L. Barnes, D.~T. Graves, T.~A. Einhorn, Fracture healing as a post-natal developmental process: Molecular, spatial, and temporal aspects of its regulation, Journal of cellular biochemistry 88~(5) (2003) 873--884.
\newblock \href {https://doi.org/10.1002/jcb.10435} {\path{doi:10.1002/jcb.10435}}.

\bibitem{barnes1999growth}
G.~L. Barnes, P.~J. Kostenuik, L.~C. Gerstenfeld, T.~A. Einhorn, Growth factor regulation of fracture repair, Journal of Bone and Mineral Research 14~(11) (1999) 1805--1815.
\newblock \href {https://doi.org/10.1359/jbmr.1999.14.11.1805} {\path{doi:10.1359/jbmr.1999.14.11.1805}}.

\bibitem{sikavitsas2001biomaterials}
V.~I. Sikavitsas, J.~S. Temenoff, A.~G. Mikos, Biomaterials and bone mechanotransduction, Biomaterials 22~(19) (2001) 2581--2593.
\newblock \href {https://doi.org/10.1016/S0142-9612(01)00002-3} {\path{doi:10.1016/S0142-9612(01)00002-3}}.

\bibitem{ferguson1999does}
C.~Ferguson, E.~Alpern, T.~Miclau, J.~A. Helms, Does adult fracture repair recapitulate embryonic skeletal formation?, Mechanisms of development 87~(1) (1999) 57--66.
\newblock \href {https://doi.org/10.1016/S0925-4773(99)00142-2} {\path{doi:10.1016/S0925-4773(99)00142-2}}.

\bibitem{vortkamp1998recapitulation}
A.~Vortkamp, S.~Pathi, G.~M. Peretti, E.~M. Caruso, D.~J. Zaleske, C.~J. Tabin, Recapitulation of signals regulating embryonic bone formation during postnatal growth and in fracture repair, Mechanisms of development 71~(1) (1998) 65--76.
\newblock \href {https://doi.org/10.1016/S0925-4773(97)00203-7} {\path{doi:10.1016/S0925-4773(97)00203-7}}.

\bibitem{reed2003vascularity}
A.~Reed, C.~Joyner, S.~Isefuku, H.~Brownlow, A.~Simpson, Vascularity in a new model of atrophic nonunion, Journal of Bone \& Joint Surgery, British Volume 85~(4) (2003) 604--610.
\newblock \href {https://doi.org/10.1302/0301-620X.85B4.12944} {\path{doi:10.1302/0301-620X.85B4.12944}}.

\bibitem{kokubu2003development}
T.~Kokubu, D.~J. Hak, S.~J. Hazelwood, A.~H. Reddi, Development of an atrophic nonunion model and comparison to a closed healing fracture in rat femur, Journal of orthopaedic research 21~(3) (2003) 503--510.
\newblock \href {https://doi.org/10.1016/S0736-0266(02)00209-7} {\path{doi:10.1016/S0736-0266(02)00209-7}}.

\bibitem{harrison2003controlled}
L.~J. Harrison, J.~L. Cunningham, L.~Str{\"o}mberg, A.~E. Goodship, Controlled induction of a pseudarthrosis: a study using a rodent model, Journal of orthopaedic trauma 17~(1) (2003) 11--21.

\bibitem{frommer1971contribution}
J.~Frommer, M.~R. Margolies, Contribution of meckel's cartilage to ossification of the mandible in mice, Journal of dental research 50~(5) (1971) 1260--1267.
\newblock \href {https://doi.org/10.1177/00220345710500052801} {\path{doi:10.1177/00220345710500052801}}.

\bibitem{shibata2014distribution}
S.~Shibata, Y.~Sakamoto, T.~Yokohama-Tamaki, G.~Murakami, B.~H. Cho, Distribution of matrix proteins in perichondrium and periosteum during the incorporation of meckel's cartilage into ossifying mandible in midterm human fetuses: An immunohistochemical study, The Anatomical Record 297~(7) (2014) 1208--1217.
\newblock \href {https://doi.org/10.1002/ar.22911} {\path{doi:10.1002/ar.22911}}.

\bibitem{cardaropoli2003dynamics}
G.~Cardaropoli, M.~Araujo, J.~Lindhe, Dynamics of bone tissue formation in tooth extraction sites, Journal of clinical periodontology 30~(9) (2003) 809--818.
\newblock \href {https://doi.org/10.1034/j.1600-051X.2003.00366.x} {\path{doi:10.1034/j.1600-051X.2003.00366.x}}.

\bibitem{araujo2005dimensional}
M.~G. Araujo, J.~Lindhe, Dimensional ridge alterations following tooth extraction. an experimental study in the dog, Journal of clinical periodontology 32~(2) (2005) 212--218.
\newblock \href {https://doi.org/10.1111/j.1600-051X.2005.00642.x} {\path{doi:10.1111/j.1600-051X.2005.00642.x}}.

\bibitem{kuboki1988time}
Y.~Kuboki, F.~Hashimoto, K.~Ishibashi, Time-dependent changes of collagen crosslinks in the socket after tooth extraction in rabbits, Journal of Dental Research 67~(6) (1988) 944--948.
\newblock \href {https://doi.org/10.1177/00220345880670061101} {\path{doi:10.1177/00220345880670061101}}.

\bibitem{simpson1960experimental}
H.~Simpson, Experimental investigation into the healing of extraction wounds in macacus rhesus monkeys., Journal of oral surgery, anesthesia, and hospital dental service 18 (1960) 391.

\bibitem{amler1969time}
M.~H. Amler, The time sequence of tissue regeneration in human extraction wounds, Oral Surgery, Oral Medicine, Oral Pathology 27~(3) (1969) 309--318.
\newblock \href {https://doi.org/10.1016/0030-4220(69)90357-0} {\path{doi:10.1016/0030-4220(69)90357-0}}.

\bibitem{christopher1941histological}
F.~R. Christopher, A histological study of bone healing in relation to the extraction of teeth, 1941.

\bibitem{davies2003understanding}
J.~E. Davies, Understanding peri-implant endosseous healing, Journal of dental education 67~(8) (2003) 932--949.

\bibitem{chandrasekhar1996modulation}
S.~Chandrasekhar, A.~K. Harvey, Modulation of pdgf mediated osteoblast chemotaxis by leukemia inhibitory factor lif, Journal of cellular physiology 169~(3) (1996) 481--490.
\newblock \href {https://doi.org/10.1002/(SICI)1097-4652(199612)169:3<481::AID-JCP8>3.0.CO;2-K} {\path{doi:10.1002/(SICI)1097-4652(199612)169:3<481::AID-JCP8>3.0.CO;2-K}}.

\bibitem{hernandez2006physiological}
I.~F.-T. Hernandez-Gil, M.~A. Gracia, M.~del Canto~Pingarrn, L.~B. Jerez, Physiological bases of bone regeneration i. histology and physiology of bone tissue, Med Oral 11 (2006) E47--51.

\bibitem{li2003plasminogen}
W.-Y. Li, S.~S. Chong, E.~Y. Huang, T.-L. Tuan, Plasminogen activator/plasmin system: a major player in wound healing?, Wound repair and regeneration 11~(4) (2003) 239--247.
\newblock \href {https://doi.org/10.1046/j.1524-475X.2003.11402.x} {\path{doi:10.1046/j.1524-475X.2003.11402.x}}.

\bibitem{aukhil2000biology}
I.~Aukhil, Biology of wound healing, Periodontology 2000 22~(1) (2000) 44--50.
\newblock \href {https://doi.org/10.1034/j.1600-0757.2000.2220104.x} {\path{doi:10.1034/j.1600-0757.2000.2220104.x}}.

\bibitem{vanegas2011finite}
J.~Vanegas-Acosta, D.~Garz{\'o}n-Alvarado, et~al., A finite element method approach for the mechanobiological modeling of the osseointegration of a dental implant, Computer Methods and Programs in Biomedicine 101~(3) (2011) 297--314.
\newblock \href {https://doi.org/10.1016/j.cmpb.2010.11.007} {\path{doi:10.1016/j.cmpb.2010.11.007}}.

\bibitem{ferrara1997biology}
N.~Ferrara, T.~Davis-Smyth, The biology of vascular endothelial growth factor, Endocrine reviews 18~(1) (1997) 4--25.

\bibitem{pugh2003regulation}
C.~W. Pugh, P.~J. Ratcliffe, Regulation of angiogenesis by hypoxia: role of the hif system, Nature medicine 9~(6) (2003) 677--684.
\newblock \href {https://doi.org/10.1038/nm0603-677} {\path{doi:10.1038/nm0603-677}}.

\bibitem{maes2012hypoxia}
C.~Maes, G.~Carmeliet, E.~Schipani, Hypoxia-driven pathways in bone development, regeneration and disease, Nature Reviews Rheumatology 8~(6) (2012) 358--366.
\newblock \href {https://doi.org/10.1038/nrrheum.2012.36} {\path{doi:10.1038/nrrheum.2012.36}}.

\bibitem{komatsu2004activation}
D.~Komatsu, M.~Hadjiargyrou, Activation of the transcription factor hif-1 and its target genes, vegf, ho-1, inos, during fracture repair, Bone 34~(4) (2004) 680--688.
\newblock \href {https://doi.org/10.1016/j.bone.2003.12.024} {\path{doi:10.1016/j.bone.2003.12.024}}.

\bibitem{hirao2006oxygen}
M.~Hirao, N.~Tamai, N.~Tsumaki, H.~Yoshikawa, A.~Myoui, Oxygen tension regulates chondrocyte differentiation and function during endochondral ossification, Journal of Biological Chemistry 281~(41) (2006) 31079--31092.
\newblock \href {https://doi.org/10.1074/jbc.M602296200} {\path{doi:10.1074/jbc.M602296200}}.

\bibitem{long2012building}
F.~Long, Building strong bones: molecular regulation of the osteoblast lineage, Nature reviews Molecular cell biology 13~(1) (2012) 27--38.
\newblock \href {https://doi.org/10.1038/nrm3254} {\path{doi:10.1038/nrm3254}}.

\bibitem{anselme2000osteoblast}
K.~Anselme, Osteoblast adhesion on biomaterials, Biomaterials 21~(7) (2000) 667--681.
\newblock \href {https://doi.org/10.1016/S0142-9612(99)00242-2} {\path{doi:10.1016/S0142-9612(99)00242-2}}.

\bibitem{olsen1997mathematical}
L.~Olsen, J.~A. Sherratt, P.~K. Maini, F.~Arnold, A mathematical model for the capillary endothelial cell-extracellular matrix interactions in wound-healing angiogenesis, Mathematical Medicine and Biology 14~(4) (1997) 261--281.
\newblock \href {https://doi.org/10.1093/imammb/14.4.261} {\path{doi:10.1093/imammb/14.4.261}}.

\bibitem{carlier2015oxygen}
A.~Carlier, L.~Geris, N.~van Gastel, G.~Carmeliet, H.~Van~Oosterwyck, Oxygen as a critical determinant of bone fracture healing - a multiscale model, Journal of theoretical biology 365 (2015) 247--264.
\newblock \href {https://doi.org/10.1016/j.jtbi.2014.10.012} {\path{doi:10.1016/j.jtbi.2014.10.012}}.

\bibitem{strehl2010flux}
R.~Strehl, A.~Sokolov, D.~Kuzmin, S.~Turek, A flux-corrected finite element method for chemotaxis problems, Computational Methods in Applied Mathematics Comput. Methods Appl. Math. 10~(2) (2010) 219--232.
\newblock \href {https://doi.org/10.2478/cmam-2010-0013} {\path{doi:10.2478/cmam-2010-0013}}.

\bibitem{strehl2013positivity}
R.~Strehl, A.~Sokolov, D.~Kuzmin, D.~Horstmann, S.~Turek, A positivity-preserving finite element method for chemotaxis problems in 3d, Journal of Computational and Applied Mathematics 239 (2013) 290--303.
\newblock \href {https://doi.org/10.1016/j.cam.2012.09.041} {\path{doi:10.1016/j.cam.2012.09.041}}.

\bibitem{brighton1972oxygen}
C.~T. BRIGHTON, A.~G. KREBS, Oxygen tension of healing fractures in the rabbit, The Journal of Bone \& Joint Surgery 54~(2) (1972) 323--332.

\bibitem{grayson2007hypoxia}
W.~L. Grayson, F.~Zhao, B.~Bunnell, T.~Ma, Hypoxia enhances proliferation and tissue formation of human mesenchymal stem cells, Biochemical and biophysical research communications 358~(3) (2007) 948--953.
\newblock \href {https://doi.org/10.1016/j.bbrc.2007.05.054} {\path{doi:10.1016/j.bbrc.2007.05.054}}.

\bibitem{chae2001hypoxia}
H.-J. Chae, S.-C. Kim, K.-S. Han, S.-W. Chae, N.-H. An, H.-M. Kim, H.-H. Kim, Z.-H. Lee, H.-R. Kim, Hypoxia induces apoptosis by caspase activation accompanying cytochrome c release from mitochondria in mc3t3e1 osteoblasts. p38 mapk is related in hypoxia-induced apoptosis, Immunopharmacology and immunotoxicology 23~(2) (2001) 133--152.
\newblock \href {https://doi.org/10.1081/IPH-100103855} {\path{doi:10.1081/IPH-100103855}}.

\bibitem{fraisl2009regulation}
P.~Fraisl, M.~Mazzone, T.~Schmidt, P.~Carmeliet, Regulation of angiogenesis by oxygen and metabolism, Developmental cell 16~(2) (2009) 167--179.
\newblock \href {https://doi.org/10.1016/j.devcel.2009.01.003} {\path{doi:10.1016/j.devcel.2009.01.003}}.

\bibitem{akasaka2010mechanisms}
Y.~Akasaka, I.~Ono, T.~Kamiya, Y.~Ishikawa, T.~Kinoshita, S.~Ishiguro, T.~Yokoo, R.~Imaizumi, N.~Inomata, K.~Fujita, et~al., The mechanisms underlying fibroblast apoptosis regulated by growth factors during wound healing, The Journal of pathology 221~(3) (2010) 285--299.
\newblock \href {https://doi.org/10.1002/path.2710} {\path{doi:10.1002/path.2710}}.

\bibitem{schweigerer1987capillary}
L.~Schweigerer, G.~Neufeld, J.~Friedman, J.~A. Abraham, J.~C. Fiddes, D.~Gospodarowicz, Capillary endothelial cells express basic fibroblast growth factor, a mitogen that promotes their own growth, Nature 325~(6101) (1987) 257--259.
\newblock \href {https://doi.org/10.1038/325257a0} {\path{doi:10.1038/325257a0}}.

\bibitem{vlodavsky1987endothelial}
I.~Vlodavsky, J.~Folkman, R.~Sullivan, R.~Fridman, R.~Ishai-Michaeli, J.~Sasse, M.~Klagsbrun, Endothelial cell-derived basic fibroblast growth factor: synthesis and deposition into subendothelial extracellular matrix, Proceedings of the National Academy of Sciences 84~(8) (1987) 2292--2296.

\bibitem{seghezzi1998fibroblast}
G.~Seghezzi, S.~Patel, C.~J. Ren, A.~Gualandris, G.~Pintucci, E.~S. Robbins, R.~L. Shapiro, A.~C. Galloway, D.~B. Rifkin, P.~Mignatti, Fibroblast growth factor-2 (fgf-2) induces vascular endothelial growth factor (vegf) expression in the endothelial cells of forming capillaries: an autocrine mechanism contributing to angiogenesis, The Journal of cell biology 141~(7) (1998) 1659--1673.
\newblock \href {https://doi.org/10.1083/jcb.141.7.1659} {\path{doi:10.1083/jcb.141.7.1659}}.

\bibitem{mehta2008platelet}
S.~Mehta, J.~T. Watson, Platelet rich concentrate: basic science and current clinical applications, Journal of orthopaedic trauma 22~(6) (2008) 432--438.
\newblock \href {https://doi.org/10.1097/BOT.0b013e31817e793f} {\path{doi:10.1097/BOT.0b013e31817e793f}}.

\bibitem{marx2001platelet}
R.~E. Marx, Platelet-rich plasma (prp): what is prp and what is not prp?, Implant dentistry 10~(4) (2001) 225--228.

\bibitem{street2002vascular}
J.~Street, M.~Bao, S.~Bunting, F.~V. Peale, N.~Ferrara, H.~Steinmetz, J.~Hoeffel, J.~L. Cleland, A.~Daugherty, N.~van Bruggen, et~al., Vascular endothelial growth factor stimulates bone repair by promoting angiogenesis and bone turnover, Proceedings of the National Academy of Sciences 99~(15) (2002) 9656--9661.
\newblock \href {https://doi.org/10.1073/pnas.152324099} {\path{doi:10.1073/pnas.152324099}}.

\bibitem{mayer2005vascular}
H.~Mayer, H.~Bertram, W.~Lindenmaier, T.~Korff, H.~Weber, H.~Weich, Vascular endothelial growth factor (vegf-a) expression in human mesenchymal stem cells: Autocrine and paracrine role on osteoblastic and endothelial differentiation, Journal of cellular biochemistry 95~(4) (2005) 827--839.
\newblock \href {https://doi.org/10.1002/jcb.20462} {\path{doi:10.1002/jcb.20462}}.

\bibitem{rupnick1988quantitative}
M.~Rupnick, C.~Stokes, S.~Williams, D.~Lauffenburger, Quantitative analysis of random motility of human microvessel endothelial cells using a linear under-agarose assay., Laboratory investigation; a journal of technical methods and pathology 59~(3) (1988) 363--372.

\bibitem{olsen1995mechanochemical}
L.~Olsen, J.~A. Sherratt, P.~K. Maini, A mechanochemical model for adult dermal wound contraction: on the permanence of the contracted tissue displacement profile, Journal of theoretical biology 177~(2) (1995) 113--128.

\bibitem{mcculloch1987paravascular}
C.~McCulloch, E.~Nemeth, B.~Lowenberg, A.~Melcher, Paravascular cells in endosteal spaces of alveolar bone contribute to periodontal ligament cell populations, The Anatomical record 219~(3) (1987) 233--242.
\newblock \href {https://doi.org/10.1002/ar.1092190304} {\path{doi:10.1002/ar.1092190304}}.

\bibitem{mcculloch1983cell}
C.~McCulloch, A.~Melcher, Cell density and cell generation in the periodontal ligament of mice, American Journal of Anatomy 167~(1) (1983) 43--58.
\newblock \href {https://doi.org/10.1002/aja.1001670105} {\path{doi:10.1002/aja.1001670105}}.

\bibitem{kingslanding}
V.~Kingsmill, C.~Gray, D.~Moles, A.~Boyde, Cortical vascular canals in human mandible and other bones, Journal of dental research 86~(4) (2007) 368--372.
\newblock \href {https://doi.org/10.1177/154405910708600413} {\path{doi:10.1177/154405910708600413}}.

\bibitem{young2008microcomputed}
S.~Young, J.~D. Kretlow, C.~Nguyen, A.~G. Bashoura, L.~S. Baggett, J.~A. Jansen, M.~Wong, A.~G. Mikos, Microcomputed tomography characterization of neovascularization in bone tissue engineering applications, Tissue Engineering Part B: Reviews 14~(3) (2008) 295--306.
\newblock \href {https://doi.org/10.1089/ten.teb.2008.0153} {\path{doi:10.1089/ten.teb.2008.0153}}.

\bibitem{chevrier2007chitosan}
A.~Chevrier, C.~Hoemann, J.~Sun, M.~Buschmann, Chitosan--glycerol phosphate/blood implants increase cell recruitment, transient vascularization and subchondral bone remodeling in drilled cartilage defects, Osteoarthritis and Cartilage 15~(3) (2007) 316--327.
\newblock \href {https://doi.org/10.1016/j.joca.2006.08.007} {\path{doi:10.1016/j.joca.2006.08.007}}.

\bibitem{lin1994differentiation}
W.-L. Lin, C.~A. McCulloch, M.-I. Cho, Differentiation of periodontal ligament fibroblasts into osteoblasts during socket healing after tooth extraction in the rat, The Anatomical Record 240~(4) (1994) 492--506.
\newblock \href {https://doi.org/10.1002/ar.1092400407} {\path{doi:10.1002/ar.1092400407}}.

\bibitem{hee2006influence}
C.~K. Hee, M.~A. Jonikas, S.~B. Nicoll, Influence of three-dimensional scaffold on the expression of osteogenic differentiation markers by human dermal fibroblasts, Biomaterials 27~(6) (2006) 875--884.
\newblock \href {https://doi.org/10.1016/j.biomaterials.2005.07.004} {\path{doi:10.1016/j.biomaterials.2005.07.004}}.

\bibitem{meikle1992human}
M.~C. Meikle, S.~Bord, R.~M. Hembry, J.~Compston, P.~I. Croucher, J.~J. Reynolds, Human osteoblasts in culture synthesize collagenase and other matrix metalloproteinases in response to osteotropic hormones and cytokines, Journal of cell science 103~(4) (1992) 1093--1099.

\bibitem{prokharau2014mathematical}
P.~A. Prokharau, F.~J. Vermolen, J.~M. Garc{\'\i}a-Aznar, A mathematical model for cell differentiation, as an evolutionary and regulated process, Computer methods in biomechanics and biomedical engineering 17~(10) (2014) 1051--1070.
\newblock \href {https://doi.org/10.1080/10255842.2012.736503} {\path{doi:10.1080/10255842.2012.736503}}.

\bibitem{gefen2011cellular}
A.~Gefen, Cellular and Biomolecular Mechanics and Mechanobiology, Springer, 2011.
\newblock \href {https://doi.org/10.1007/978-3-642-14218-5} {\path{doi:10.1007/978-3-642-14218-5}}.

\bibitem{thompson2002model}
Z.~Thompson, T.~Miclau, D.~Hu, J.~A. Helms, A model for intramembranous ossification during fracture healing, Journal of Orthopaedic Research 20~(5) (2002) 1091--1098.
\newblock \href {https://doi.org/10.1016/S0736-0266(02)00017-7} {\path{doi:10.1016/S0736-0266(02)00017-7}}.

\bibitem{prendergast2010computational}
P.~Prendergast, S.~Checa, D.~Lacroix, Computational models of tissue differentiation, in: Computational modeling in biomechanics, Springer, 2010, pp. 353--372.
\newblock \href {https://doi.org/10.1007/978-90-481-3575-2_12} {\path{doi:10.1007/978-90-481-3575-2_12}}.

\end{thebibliography}

\appendix

\section{Non-dimensional parameters} \label{appendixA}
Equations \ref{eq:cm}-\ref{eq:gv} were non-dimensionalized as proposed by Geris \textit{et al.} \cite{geris2008angiogenesis}, but as mentioned before, we made some modifications, in order improve the robustness of the FEM implementation. The scalings are the following:\\
\begin{small}
\begin{equation}
\begin{split}
\tilde t = \frac{t}{T},\quad \tilde x_h  = \frac{{x_h }}{L},\quad \tilde c_i  = \frac{{c_i }}{{c_0 }},\quad \tilde c_v  = \frac{{c_v }}{{c_{v,0} }}, \\ \quad \tilde m_j  = \frac{{m_j }}{{m_0 }},\quad \tilde m_v  = \frac{{m_v }}{{m_{v,0} }},\quad \tilde g_k  = \frac{{g_k }}{{g_0 }}
\end{split}
\end{equation}\\
\end{small}
Where $x_h$ are the coordinates that define the position of a point ($r$ and $z$ for the axisymmetric model); $i=m,f,b$; $j=m,b$ and $k=b,v$. The characteristic scales for the variables were chosen exactly the same as in \cite{geris2008angiogenesis}, except for $c_{v,0}$ (which was changed to avoid great differences of order of magnitude), $m_{v,0}$ (which was changed automatically when adopting the vascularization model of Olsen \textit{et al.} \cite{olsen1997mathematical}) and $L$ (which was chosen using optimization algorithm). The characteristic scales are $T=1$ day, $L=1$ mm, $m_0=0.1$ g/ml, $g_0=100$ ng/ml, $c_0=10^6$ cells/ml and $c_{v,0}=10^2$ cells/ml. Since $T=1$ mm, the alveolar geometry can be modeled in mm and it is not necessary to scale it.\\

\begin{table*}[h] 
  \centering
  \caption{Adimensional parameters obtained for the calculations of this study (excluding differentiation parameters - Table \ref{tab:diffparam}). The sub-index $x$ of the parameters in the first column is to be replaced with the letters of the first row. Calculation of $D_x$; $C_{x,CT}$; $C_{x,HT}$; and $A_x$ with values of first 8 rows is explained in \cite{geris2008angiogenesis}}
  \begin{footnotesize}   
   
    \begin{tabular}{rrrrrp{5.5cm}}
    \hline
          & m     & f     & b     & v     & Description \\
    \hline
    \multicolumn{6}{l}{\textit{\textbf{For cells}}} \\
    $\tilde D_{hx}$ & 0.1715 & 0.00147* &       & 0.1417 & Diffusivity parameter 1  \\
    $\tilde K_{hx}$ & 0.25  &       &       & 1     & Diffusivity parameter 2  \\
    $\tilde C_{kCTx}$ & 0.49  & 4.9   &       & 0.37485 & Chemotaxis parameter 1  \\
    $\tilde K_{kCTx}$ & 0.1   & 0.1   &       & 0.025 & Chemotaxis parameter 2  \\
    $\tilde C_{kHTx}$ & 0.04165 &       &       & 0.03** & Haptotaxis parameter 1  \\
    $\tilde K_{kHTx}$ & 0.5   &       &       &       & Haptotaxis parameter 2  \\
    $\tilde A_{x0}$ & 1.01  & 0.202 & 0.202 & 0.202 & Proliferation rate param. 1  \\
    $\tilde K_x$ & 0.1   & 0.1   & 0.1   & 0.1   & Proliferation rate param. 1  \\
    $\tilde \alpha_x$ & 1     & 1     & 1     & 1 & Saturation for proliferation \\
    $\tilde d_{x,constant}$ &       &       &       & 0.1   & Constant apoptosis rate \\
    $\tilde d_{x,hyp}$ & 10    & 10    & 10    &       & H-D apoptosis rate \\
    $\tilde H_{hyp,x}$ & 0.04167 & 0.1875 & 0.167 &       & Threshold for H-D apoptosis \\
    $\tilde d_{x,bFGF}$ &       & 0.45     &       &       & bFGF-dependent apoptosis rate \\
    $\tilde H_{bFGF,x}$ &       & 0.92  &       &       & Threshold for bFGF-dependent apoptosis \\
    \multicolumn{6}{l}{\textit{\textbf{For extra-cellular matrices}}} \\
    $\tilde P_{xs}$ &       & 0.2   & 2     & 1   & Synthesis rate \\
    $\tilde \kappa_x$ &       & 1     & 1     & 1     & Maximum matrix density \\
    $\tilde Q_x$ &       & 1.5   &       &       & Resorption rate \\
    \multicolumn{6}{l}{\textit{\textbf{For growth factors}}} \\
    $\tilde D_{gx}$ &       &       & 0.06125 & 6.125 & Diffusivity \\
    $\tilde G_{gbx}$ &       &       & 1000  &       & $\tilde g_b$ production rate \\
    $\tilde G_{gvx}$ &       &       & $5\times 10^{-5}$ &       & $\tilde g_v$ production rate \\
    $\tilde H_{gx}$ &       &       & 1     & 0.03     & Threshold for growth factor production \\
    $\tilde d_{gx}$ &       &       & 100   & 30    & Decay rate \\
    $\tilde d_{gvx}$ &       &       &       & 20    & Consumption rate of $\tilde g_v$ \\
    $\tilde \epsilon_{hyp,x}$ & 1     & 1     & 1     &       & H-D $\tilde g_v$ production rate \\
    $\tilde K_{hyp,x}$ & 0.167 & 0.375 & 0.33  &       & Threshold for H-D $\tilde g_v$ production \\
$\tilde \epsilon_{clot}$ &      &  30     &      &       & $\tilde g_v$ production rate by BC\\
    $\tilde K_{clot}$ & & 0.350 &  &       & Threshold for  $\tilde g_v$ production by BC\\    
    \hline
    \multicolumn{6}{l}{*Value for constant diffusivity of fibroblasts (ignore $\tilde K_{hf}$)} \\
    \multicolumn{6}{l}{**Value for constant haptotaxis parameter of endothelial cells (ignore $\tilde K_{v,HT}$)} \\
    \end{tabular}%
   \end{footnotesize}
  \label{tab:params}%
\end{table*}%

\begin{table*}[h] 
  \centering
  \caption{Non-dimensional differantiation parameters used for the simulations.}
    \begin{tabular}{ccl}
    \hline
    Parameter & Value & Description \\
    \hline
    $\tilde Y_{11}$ & 20 & MSC to osteoblast differentiation rate 1 (see \cite{geris2008angiogenesis})\\
    $\tilde H_{11}$ & 0.1 & MSC to osteoblast differentiation parameter 1 (see \cite{geris2008angiogenesis})\\
    $\tilde Y_{12}$ & 2.3 & MSC to osteoblast differentiation rate 2 (see \cite{geris2008angiogenesis})\\
    $\tilde H_{12}$ & 0.1 & MSC to osteoblast differentiation parameter 2 (see \cite{geris2008angiogenesis})\\
    $\tilde I_v$ & 0.667 & Minimum vascularization needed for $c_m \rightarrow c_b$ differentiation\\
    $\tilde F_{mf}$ & 0.01 & MSC to fibroblast differentiation rate\\
    $\tilde F_{bo}$ & 0.01 & Osteoblast to osteocyte differentiation rate\\
    \hline
    \end{tabular}%
  \label{tab:diffparam}%
\end{table*}%
\pagebreak

\end{document}